\documentclass[fleqn,10pt]{wlscirep}
\usepackage[utf8]{inputenc}
\usepackage[T1]{fontenc}

\usepackage{multirow, bigstrut}
\usepackage[normalem]{ulem}
\usepackage{textcomp}

\usepackage{xcolor}

\usepackage{enumitem}
\setlist{nosep}

\newif\ifexportfigs
\ifexportfigs
    \usepackage[floats,active,tightpage]{preview}
\fi

\title{Free-moving Quantitative Gamma-ray Imaging}
\author[1,*]{Daniel~Hellfeld}
\author[1]{Mark~S.~Bandstra}
\author[1]{Jayson~R.~Vavrek}
\author[2]{Donald~L.~Gunter}
\author[1]{Joseph~C.~Curtis}
\author[1]{Marco~Salathe}
\author[1]{Ryan~Pavlovsky}
\author[1]{Victor~Negut}
\author[1]{Paul~J.~Barton}
\author[1]{Joshua~W.~Cates}
\author[1]{Brian~J.~Quiter}
\author[1]{Reynold~J.~Cooper}
\author[1, 3]{Kai~Vetter}
\author[1]{Tenzing~H.\,Y.~Joshi}
\affil[1]{Nuclear Science Division, Lawrence Berkeley National Laboratory, Berkeley, CA 94720 USA}
\affil[2]{Gunter Physics, Inc., Lisle, IL 60532 USA}
\affil[3]{Department of Nuclear Engineering, University of California, Berkeley, Berkeley, CA 94720 USA}
\affil[*]{To whom correspondence should be addressed; E-mail: dhellfeld@lbl.gov.}

\begin{abstract}
The ability to map and estimate the activity of radiological source distributions in unknown three-dimensional environments has applications in the prevention and response to radiological accidents or threats as well as the enforcement and verification of international nuclear non-proliferation agreements.
Such a capability requires well-characterized detector response functions, accurate time-dependent detector position and orientation data, a digitized representation of the surrounding 3D environment, and appropriate image reconstruction and uncertainty quantification methods.
We have previously demonstrated 3D mapping of gamma-ray emitters with free-moving detector systems   on a relative intensity scale using a technique called Scene Data Fusion (SDF).
Here we characterize the detector response of a multi-element gamma-ray imaging system using experimentally benchmarked Monte Carlo simulations and perform 3D mapping on an absolute intensity scale.
We present experimental reconstruction results from hand-carried and airborne measurements with point-like and distributed sources in known configurations, demonstrating \textit{quantitative} SDF in complex 3D environments.
\end{abstract}
\begin{document}

\flushbottom
\maketitle
\thispagestyle{empty}

\section*{Introduction}

The reconstruction of the absolute intensity distribution of radioactive materials in complex 3D scenes has numerous applications including geologic survey, contamination assessment, guidance for decontamination and remediation verification, nuclear facility decommissioning, radiation safety/protection, and nuclear security.
For example, 3D distributed source mapping is critical in response to radiological contamination such as in the Fukushima Prefecture in Japan, Chernobyl in Ukraine, or more recently the University of Washington Harborview Training and Research Facility where a sealed 2.9 kCi \textsuperscript{137}Cs source was inadvertently breached \cite{jit2020}.
In these cases, the assessment of the strength and distribution of the contamination was and remains to be necessary to minimize operator risk as well as to enable more effective means in the communication and visualization of nuclear radiation with the public to inform evacuation, decontamination, resettlement, and reopening efforts \cite{vetter_2019}.
Additionally, such a capability can aid in consequence management efforts in order to avoid exposure to or plan dose-minimizing paths through radioactive fallout \cite{mcmanus_2019}.
Furthermore, this capability facilitates search applications of compact radiological sources such as lost sources \cite{aage_2003}, \emph{broken arrows} (i.e., accidents involving nuclear weapons and components) \cite{croddy_2018}, as well as specific missions of the U.S.~Special Operations Command such as sensitive site exploitation and nuclear disablement \cite{army2015,mcmanus_2019}.

The work presented here is part of an effort undertaken at Lawrence Berkeley National Laboratory (LBNL) and the University of California, Berkeley to develop a robust framework for performing 3D radiological imaging using platform-agnostic and free-moving detection and imaging systems.
The emphasis of this paper is placed on quantitative gamma-ray image reconstruction, in that 3D gamma-ray maps are reconstructed accurately with relevant units such as activity of an isotope of interest (in \textmu Ci or MBq).
In principle, the activity maps can also be used to compute 3D dose-rate maps (in \textmu Sv$\cdot$hr$^{-1}$) relevant for contamination assessment, emergency response, and radiation protection.
Quantitative assessments are performed across scenes and environments (i.e., not just at the location of the instrument) and systems can be deployed on unmanned and remotely operating platforms to minimize risk to human operators if necessary.

This paper begins by presenting the necessary background and overview of the quantitative free-moving 3D imaging approach, including data fusion methods and detector response characterization.
Experimental demonstrations for both point-like and distributed gamma-ray sources are described and the results are discussed -- highlighting the successes, current limitations, and potential areas for future work.
Details regarding the gamma-ray imaging system, detector response characterization, image reconstruction algorithms, and uncertainty quantification are provided in the Methods section.

\paragraph*{Free-moving 3D Imaging}

Compact gamma-ray imaging devices are available commercially \cite{wahl_2015,phds} and have seen wide use in the field.
For example, products sold by the Michigan-based radiation detection and imaging company H3D Inc.\,\,are deployed at $>$75\% of nuclear power plants in the U.S. \cite{h3dwebsite} and integrated into U.S. Department of Defense systems such as the MERLIN/VIPER vehicle \cite{webber_2018}.
However, these systems require static operation as they possess no inherent method of self-localization relative to a global imaging frame.
Similar to traditional visual photography, the images produced are two dimensional and are relative only to the position and orientation in which the data were collected.
If overlaid on a co-registered visual camera image, the 2D gamma-ray image can provide the user some context as to the direction of the source relative to the scene, however, due to the inherent penetrative power of gamma-rays and the often three-dimensional nature of distributed sources (e.g., uranium storage tanks, radioactive holdup, radioactive contamination, etc.), accurate depth estimation is critical in determining the location, distribution and activity associated with sources in a 3D environment.
Given no information of the source distribution prior to imaging, the static data collection process is also limited by potentially large source standoffs and/or complex directional source shielding, thus requiring long dwell times (i.e., tens of minutes to several hours) to acquire enough data to produce spatially accurate images.
Moreover, the resultant 2D images are not quantitative, unless in specific cases where the distance to the imaged source distribution is somehow specified \cite{vetter_2018}.

The ability to move the system during the data collection process can improve detection sensitivity and significantly reduce the time to image (i.e., to a few minutes or less) by getting closer to weaker sources, viewing sources from various directions to break directional degeneracies, or potentially finding narrow streaming paths in complex shielding configurations \cite{sinclair_2020}.
Moreover, data collected at multiple known positions and orientations in a global coordinate frame can be combined to facilitate 3D image reconstruction.
Taking images from multiple perspectives is standard practice in biomedical imaging, where detector systems are rotated along a precisely known trajectory around a stationary image space in order to produce 3D gamma-ray images of administered radioactive compounds.
In the nuclear security domain, however, neither the system trajectory nor the image space are usually well-defined before a measurement begins.
For example, in the source search scenario, the path a user takes may depend on the environment they are operating in and any feedback given to them during the measurement (e.g., elevated count-rate or source detection/identification alarms).
Moreover, the user may explore various areas of interest, in which case an adaptive image space is required.

By correlating the radiation data with the globally-consistent system pose (i.e., position and orientation), the radiation data can be transformed into the global coordinate frame and combined in a consistent manner to produce a single 3D image.
Previous work building upon Vetter et al.\cite{vetter_2007} demonstrated this concept in the nuclear security domain, though it involved performing manual pose estimation using a laser rangefinder at several static measurement positions around a gamma-ray source of interest \cite{mihailescu_2009}.
LBNL has since pursued the development of an approach, referred to as nuclear Scene Data Fusion (SDF) \cite{barnowski_2015,vetter_2016,haefner_2017,pavlovsky_2018,hellfeld_2019_2,vetter_2019}, that uses Simultaneous Localization and Mapping (SLAM) \cite{durrant-whyte_2006_1,durrant-whyte_2006_2} to produce accurate and globally consistent pose estimates (i.e., the localization) and 3D models (i.e., the map) of the environment in real-time, facilitating free-moving 3D gamma-ray imaging.
Mapping can and has been demonstrated with both structured light \cite{barnowski_2015,haefner_2017} and lidar sensors \cite{pavlovsky_2018,hellfeld_2019_2}.
Similar work has also been demonstrated, including lidar-based SLAM with a mobile ground-robot \cite{kim_2017}, offline photogrammetry (in place of SLAM) for ground-based \cite{sato_2019} and airborne systems \cite{joshi_2017}, as well as object detection, tracking and source attribution in visual and lidar data for static systems \cite{henderson_2020,marshall_2020}.

With the ability to continuously locate the detector system in real-time, the imaging operation becomes platform agnostic -- meaning the system can be translated and rotated by any means (e.g., human, automobile, robot, aerial system, etc.).
Furthermore, with the ability to map the surrounding environment, an adaptive image space can be defined and continuously updated as the system explores new space.
The image space is typically voxelized at some resolution (e.g., 20~cm) and can be constrained by the 3D map provided by SLAM, meaning only voxels which represent occupied space are considered in the image reconstruction.
In the case of lidar-based SLAM, the 3D map is represented as a point cloud (i.e., a collection of lidar returns oriented in 3D space) and voxels are considered occupied if they contain one or a configurable number of points.
The occupancy constraint forces the reconstructed activity to physical objects in the scene in which the lidar measured a reflection, generally improving image accuracy and decreasing noise (i.e., spurious activity) under the assumption that gamma-ray sources are not present in free space.
A consequence of the current implementation is that voxels inside of enclosed objects are considered free and thus ignored, limiting the approach to source localization and emission rate reconstruction on the surfaces of objects.
The constraint also tends to reduce the number of reconstructed voxels by over a factor of 10, decreasing typical reconstruction times to the order of seconds when GPU-parallelized.
Numerous 3D image estimates can be returned and displayed to the user during the course of a free-moving measurement, allowing for course correction or detailed survey of potential source locations.

The SDF approach is multi-modal and has been demonstrated with proximity mapping \cite{pavlovsky_2018} as well as active coded mask \cite{hellfeld_2019_2}, using photo-electric absorption-only or ``singles'' events, and Compton imaging \cite{barnowski_2015,haefner_2017}, using two-site interactions (Compton scatter and photo-electric absorption) or ``doubles'' events.
In addition to gamma-rays, SDF has also demonstrated 3D mapping of neutron sources \cite{pavlovsky_2019_2}.
Moreover, to specifically reconstruct point-sources, we have developed an iterative approach called Point-Source Localization (PSL) capable of localizing one or multiple point-sources in a discrete or continuous image space \cite{hellfeld_2019_1,vavrek_2020}.

In this work, we demonstrate near real-time accurate quantitative free-moving gamma-ray imaging with uncertainty estimation.
This is demonstrated using four different imaging modalities: a refined version of discrete PSL with (a) singles and (b) Compton events, and Maximum \emph{A Posteriori} Expectation Maximization (MAP-EM) that includes a sparsity constraint ($L_{1/2}$ norm \cite{qian_2011}) with (c) singles and (d) Compton events.
These four methods provide complementary coverage over the possible application space, spanning low (singles) and high (Compton) energy emissions and both point (PSL) and distributed (MAP-EM) source scenarios.
Further detail of the reconstruction algorithms developed and used in this work can be found in the Methods section.

\paragraph*{Quantitative Reconstruction}

Prior work with the SDF approach focused primarily on the development of the free-moving 3D imaging methods for various contextual and radiation sensor combinations and imaging modalities.
The detector response functions used in reconstructions were not simulated or measured with respect to an absolute intensity scale and thus the gamma-ray activity estimates associated with the reconstructed images were qualitative in nature.
The units associated with a gamma-ray image are critical to provide actionable information to operational personnel in applications such as contamination assessment, remediation verification and nuclear safeguards.
Additionally, absolute units are necessary in biomedical imaging to provide quantitative assessment of disease and/or effectiveness of radiation-based treatments and therapies.
In this work, we characterize the detector response functions with experimentally benchmarked Geant4~\cite{agostinelli_2003,allison_2006,allison_2016} simulations in order to produce quantitative images (see Methods).

\paragraph*{Experimental Demonstrations}

Two source scenarios are considered and presented here.
As an example of a port security application, the first scenario demonstrates the localization and quantification of a single point-source in a wide-area outdoor environment involving stacks of cargo containers, using a small Unmanned Aerial System (sUAS)-based operation to move the radiation detection payload.
The second scenario demonstrates the reconstruction of multiple distributed sources in a cluttered indoor laboratory environment, modeling a radioactive spill that can be surveyed using a detector system in a hand-held operation.

\begin{figure}[htb!]
    \begin{center}
    \begin{tabular}{cc}
        \multicolumn{1}{l}{\bf{a}} & \multicolumn{1}{l}{\bf{b}} \\[2pt]
        \fbox{\includegraphics[height=0.36\columnwidth]{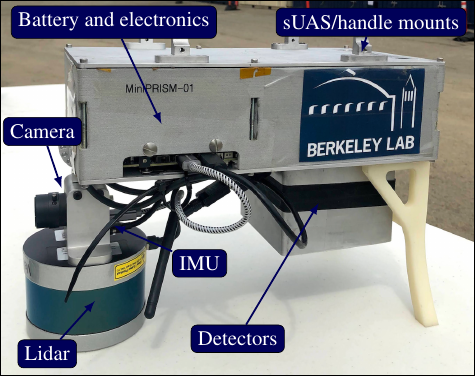}} &
        \includegraphics[height=0.33\columnwidth]{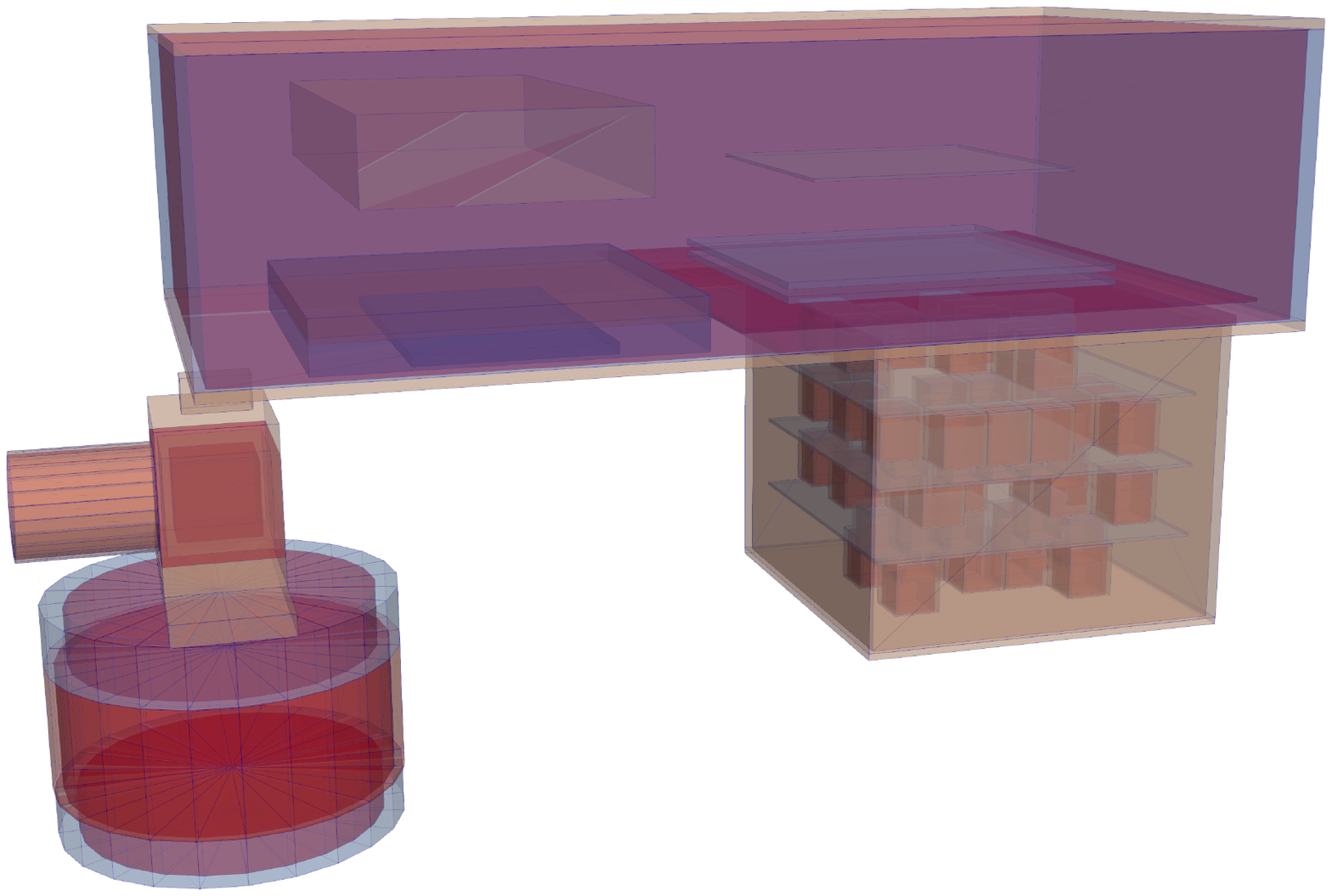}
    \end{tabular}
    \end{center}
    \vspace{-10pt}
    \ifexportfigs
    \else
        \caption{{\bf Free-moving gamma-ray imaging system}. ({\bf a}) The MiniPRISM system shown with individual components labeled. ({\bf b}) Rendering of the Geant4 simulation model geometry.}
    \fi
    \label{fig:miniprism_system}
\end{figure}

The SDF detector platform used in this work is MiniPRISM \cite{pavlovsky_2019_1} -- a $6 \times 6 \times 4$ array partially populated with 1-cm$^3$ coplanar-grid CdZnTe (CZT) detectors \cite{luke_1994,luke_1995} combined with a contextual sensor suite and on-board computer to perform lidar-based SLAM with Google Cartographer \cite{hess_2016}.
At the time of this measurement, a total of 58 detectors were loaded in the array in an optimized configuration to balance detection sensitivity and active coded mask imaging performance \cite{hellfeld_2017}.
The MiniPRISM detector system and a rendering of the simulation model are shown in Fig.~\ref{fig:miniprism_system}.

\section*{Results}

\subsection*{Point Source Localization}

\begin{figure*}[htb!]
    \centering
    \renewcommand{\arraystretch}{1}
    \setlength{\tabcolsep}{3pt}
    \begin{tabular}{cc}
        \multicolumn{2}{l}{\bf{a}} \\
        \multicolumn{2}{c}{\includegraphics[width=0.95\textwidth]{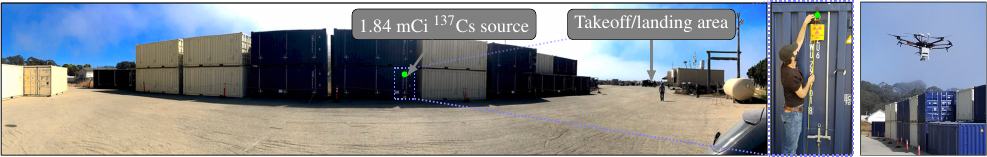}}
        \\
        \multicolumn{1}{l}{\bf{b}} &  \\[-7pt]
        \multicolumn{2}{c}{\includegraphics[width=0.88\textwidth]{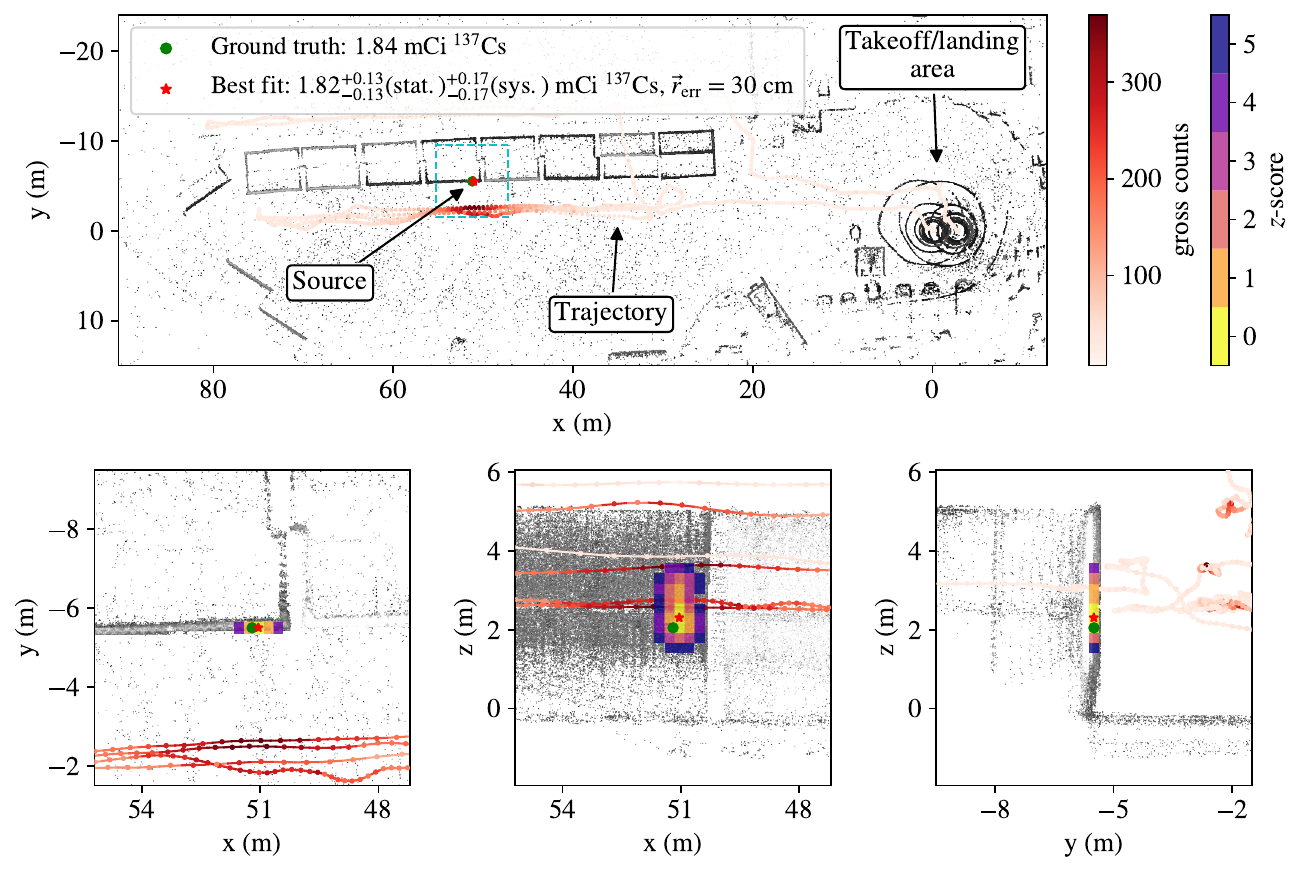}} \\[-7pt]
        \multicolumn{1}{l}{\bf{c}} & \multicolumn{1}{l}{\bf{d}} \\
        \includegraphics[height=0.26\textwidth]{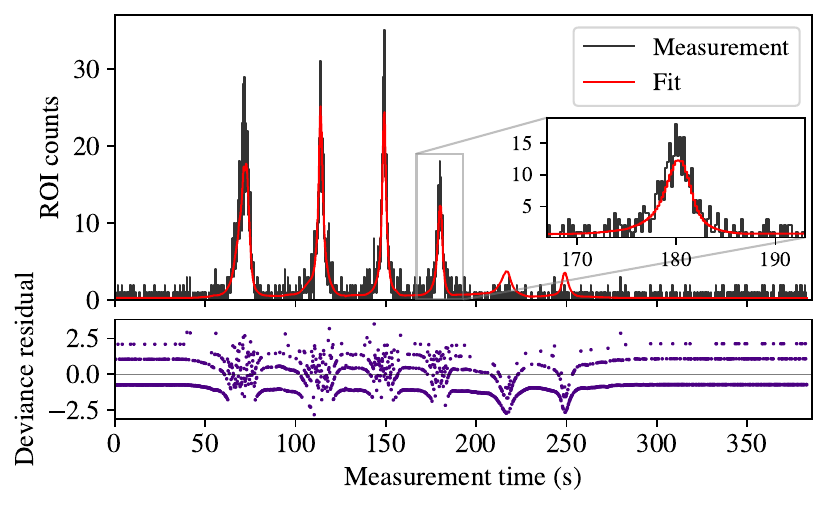} \bigstrut &
        \fbox{\includegraphics[height=0.26\textwidth]{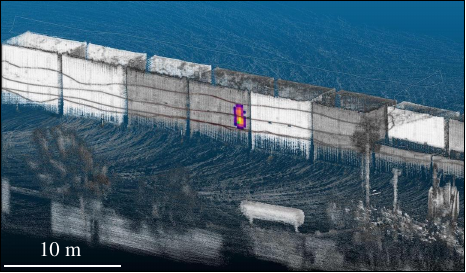}} \bigstrut \\[-4pt]
    \end{tabular}
        \ifexportfigs
    \else
    	\caption{{\bf Point-source reconstruction results}. ({\bf a}) Point-source scenario in which a 1.84~mCi \textsuperscript{137}Cs source was placed $\approx$2.5~m above the ground on the exterior of a cargo container stack. MiniPRISM was flown remotely on a sUAS and surveyed the stack in $\approx$7~min. ({\bf b}) Quantitative active coded mask PSL reconstruction following the survey of the cargo containers, including a full top-down projection as well as three zoomed-in projections near the true source location. ({\bf c}) The forward projection of the best-fit model overlaid on the measured data, summed across all detectors. The discrepancy between the measurement and model is shown with the deviance residual. ({\bf d}) Rendering of the colorized 3D point cloud.}
   \fi
    \label{fig:point_source_results}
\end{figure*}

A 1.84 mCi \textsuperscript{137}Cs point-source was placed $\approx$2.5~m above the ground on the exterior of a steel cargo container in a mock port setting with $>$20 containers, some stacked two containers high (see Fig.~\ref{fig:point_source_results}a).
The uncertainty of the source activity was undocumented, though we assume a 5--10\% uncertainty (where uncertainty here is at 95\% confidence or 2$\sigma$) based on previous purchases from the manufacturer.
The MiniPRISM system was mounted to a sUAS and flown around the containers, mapping an area of $>$4,000~m$^2$ in under 7~min.
The distance of closest approach to the source was $\approx$3~m.
The average and maximum speed of the sUAS during the measurement were 1.4 and 4.6 m$\cdot$s$^{-1}$, respectively.

In contrast to online operation where multiple reconstructions can be performed as data are collected in real-time, the results presented here were generated as a single reconstruction that includes all data over the course of the measurement.
The reconstruction energy region-of-interest (ROI) was set to $662 \pm 60$~keV to encompass the entire \textsuperscript{137}Cs photo-peak for both singles (active coded mask) and doubles (Compton) events.
While the expected energy resolution of coplanar-grid CZT is 1--3\% \cite{luke_2007}, the large energy ROI was necessary due to nonlinearities and gain drifts across individual detectors during the measurement.
Here the doubles events refer to two-interaction sequences within the detector system in a coincident time window of 36~\textmu s.
A total of 2864~singles and 370~doubles were recorded in the ROI over the entire measurement.
Other reconstruction parameters include the following: 25~cm voxels, occupancy threshold of 5 points from the point cloud, 97,310~occupied voxels (1.5\% of total), 1,916~poses (separated in time at 5~Hz) and 20~PSL iterations.
At the time of the measurement, 7 of the 58~detectors were suffering from electronics problems and the counts were excluded from the reconstruction (though the occlusion provided by the detector material was still accounted for in the detector response).

The number of iterations used in maximum likelihood problems typically represents a balance between contrast recovery and image noise amplification \cite{conti_2007}.
Statistical stopping criteria have been explored, but generally do not translate to image quality metrics in terms of noise or artifacts \cite{pafilis_2001}.
Currently the number of iterations is set based on past experience and reconstruction speed requirements.

The active coded mask PSL reconstruction results are shown in Fig.~\ref{fig:point_source_results}b where the localization estimates are displayed as Gaussian $z$-scores that represent the result of a likelihood ratio test between each voxel and the best-fit voxel (marked with a red star).
The voxelized image estimate is overlaid on the measured point cloud, colorized by the lidar return intensities, and the system trajectory, colorized by the gross counts in the detector.
A top-down projection of the entire measurement space is shown, as well as three projections zoomed in near the true source location.
Larger $z$-scores represent a larger deviation from the best fit and represent the spatial uncertainty of the best-fit estimate.
In this visualization, $z$-scores above 5$\sigma$ are clipped from the image.
The true source location (marked with a green dot) is 30~cm from the center of the best-fit voxel and is contained within the 2$\sigma$ spatial uncertainty.
The estimated activity is $1.82^{+0.13}_{-0.13}(\textrm{stat.})^{+0.17}_{-0.17}(\textrm{sys.})$~mCi of \textsuperscript{137}Cs (2$\sigma$ uncertainty bounds), and contains the true source activity of 1.84~mCi.
The statistical uncertainty is calculated in the PSL algorithm (see Methods) and the systematic uncertainty is calculated from a 9.4\% systematic uncertainty (2$\sigma$) ascribed to the detector response due to uncertainty in the calibration source activity and source-detector distance used in the detector response characterization.
Systematic uncertainty associated with the source-detector orientation during the response characterization is expected to be small, but was not quantified here.

Figure~\ref{fig:point_source_results}c shows the forward projection of the reconstructed PSL image (i.e., point-source and background) into data space overlaid on the measured data, summed across all detectors, as well as the deviance residual \cite{mccullagh_1989} between the measurement and fit.
The fit represents the maximum likelihood solution given the point-source constraint implemented in PSL.
The discrete separations observed in the deviance residuals are an artifact of low statistics and the discrete nature of the measured data (i.e., 0, 1 or 2 counts per bin).
The attenuating effect of the steel containers is not included in the physics model and thus as the system surveys the back-side of the container stack, the forward-projected signal is expected to increase when approaching the source location.
This causes two small unobserved peaks in the model (roughly 215 and 250~s into the measurement).
The inclusion of attenuation in the reconstruction is part of an ongoing effort \cite{bandstra_2021} but not included here.
Finally, Fig.~\ref{fig:point_source_results}d shows an isometric rendering of the measured point cloud, colorized by the lidar return intensities as well as the PSL $z$-scores, demonstrating the rich 3D contextual detail that can be provided via lidar-based SLAM.
While additional compute time is necessary for generating these visualizations, the reconstruction result was computed in approximately 27~s on the integrated GPU (Intel Iris Plus Graphics 650) on the on-board computer in MiniPRISM.

The Compton PSL results are similar to Fig.~\ref{fig:point_source_results} and are included in Supplementary Fig.~S1.
They also include the true source location within 2$\sigma$ spatial bounds, and the best-fit voxel to true source location distance is 9~cm.
The source activity estimate is
$2.32^{+0.42}_{-0.40}(\textrm{stat.})^{+0.22}_{-0.22}(\textrm{sys.})$~mCi of \textsuperscript{137}Cs (2$\sigma$~uncertainty bounds).
The estimate is consistent with the ground-truth activity as well as with the active coded mask PSL activity estimate.
The total reconstruction time for the Compton PSL reconstruction was 7.7~s.
The faster reconstruction time in this case is attributed to the list-mode implementation of the Compton PSL algorithm (see Methods) and the smaller number of measured doubles events.

\subsection*{Distributed Source Imaging}

Several hundred 1-\textmu Ci \textsuperscript{22}Na point-sources, certified to 5\% uncertainty by the manufacturer, were
used to create a mock distributed source with known extent and activity.
To ensure the imaging system was unable to distinguish the point-like nature of the sources, the individual sources were placed close together (5.08~cm pitch), the detector maintained at least a 50~cm standoff from the sources, and the voxelization scale of the reconstruction space was set at twice the source pitch.

\begin{figure}[htb!]
    \centering
    \setlength{\tabcolsep}{5pt}
    \begin{tabular}{*{3}{@{\hspace*{1mm}}c}}
        \multicolumn{1}{l}{{\bf a.1}} & \multicolumn{1}{l}{{\bf a.2}} & \multicolumn{1}{l}{{\bf a.3}} \\
        \fbox{\includegraphics[width=0.31\textwidth]{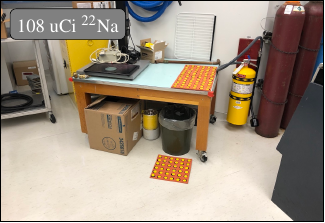}} &
        \fbox{\includegraphics[width=0.31\textwidth]{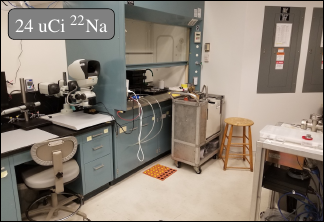}} &
        \fbox{\includegraphics[width=0.31\textwidth]{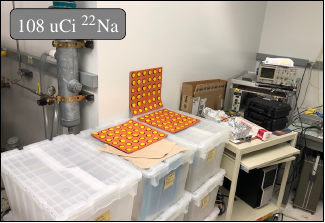}} \\

        \multicolumn{1}{l}{\bf{b.1}} & \multicolumn{1}{l}{\bf{b.2}} & \multicolumn{1}{l}{\bf{b.3}} \\[-1pt]
        \fbox{\includegraphics[width=0.31\textwidth]{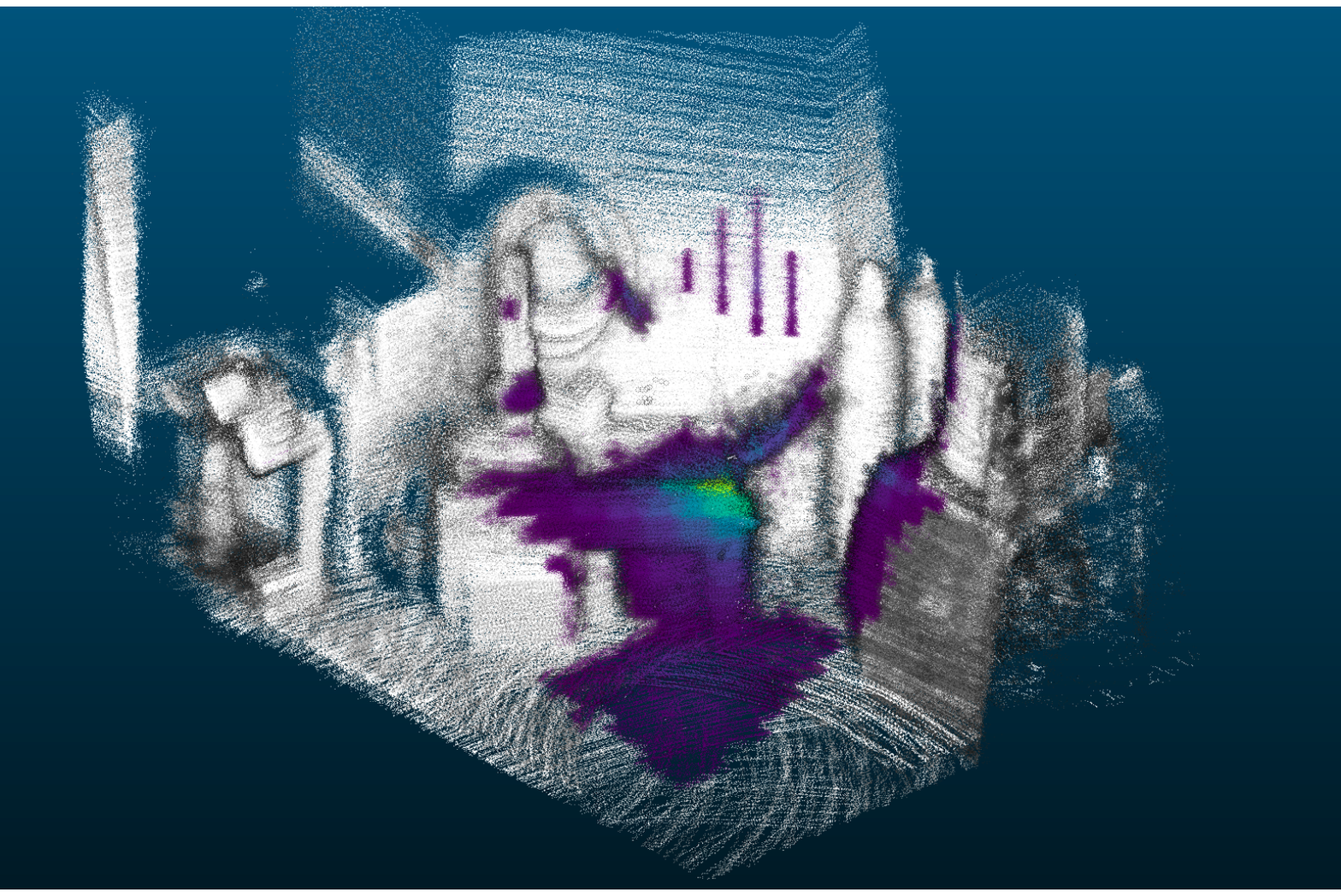}} &
        \fbox{\includegraphics[width=0.31\textwidth]{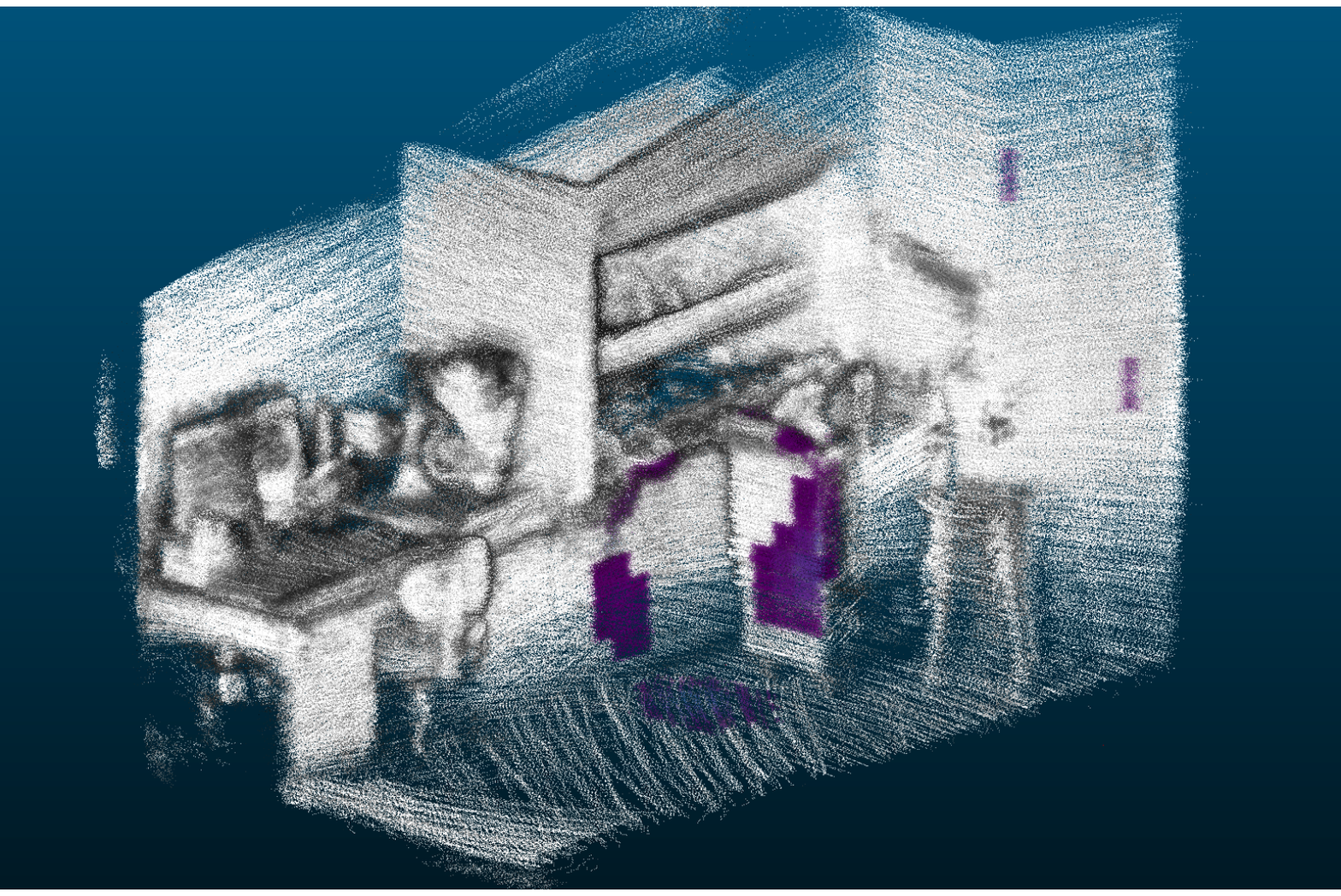}} &
        \fbox{\includegraphics[width=0.31\textwidth]{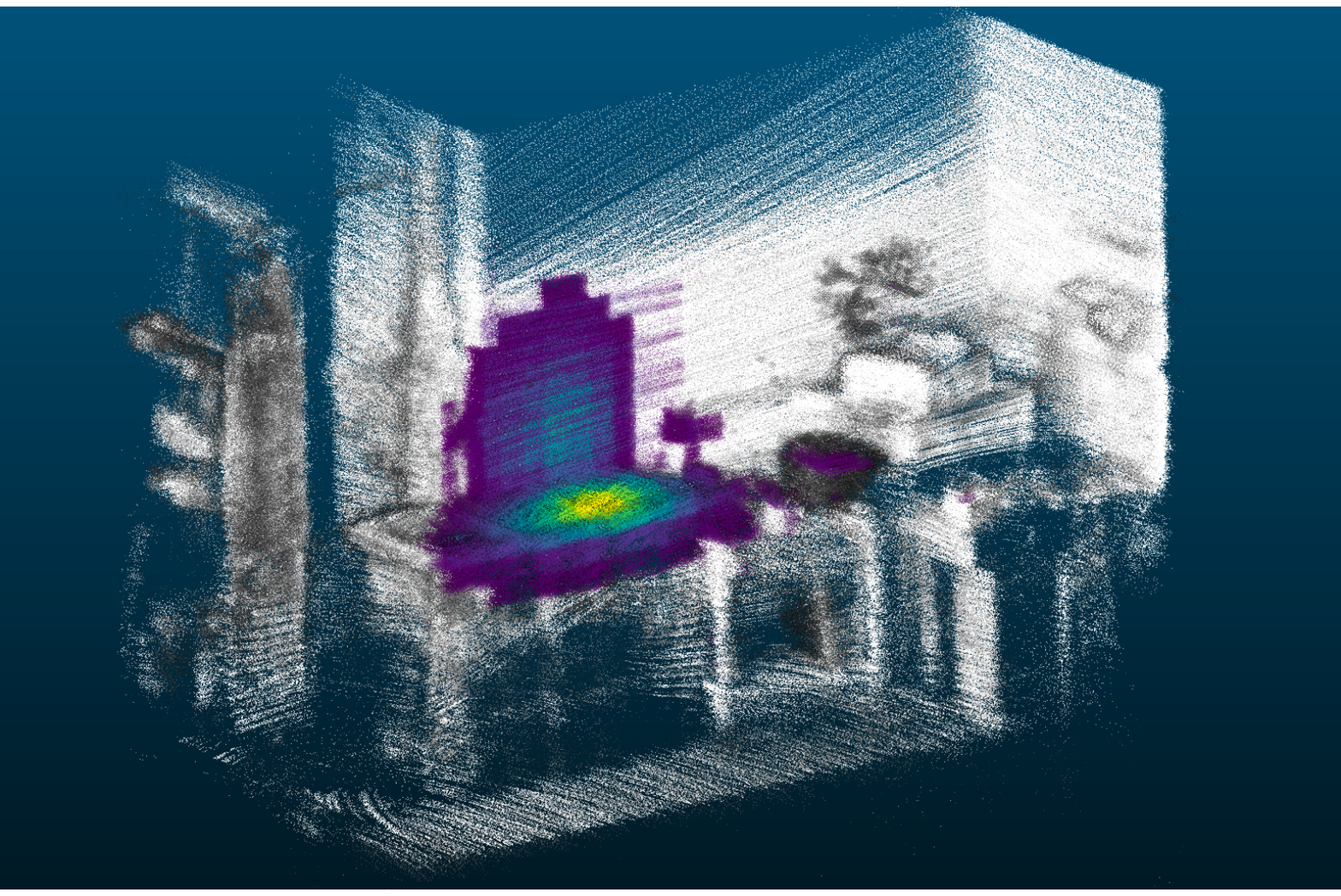}} \\

        \multicolumn{1}{l}{\bf{c.1}} & \multicolumn{1}{l}{\bf{c.2}} & \multicolumn{1}{l}{\bf{c.3}} \\[-3pt]
        \hspace*{-0.3cm}\includegraphics[height=0.3\textwidth]{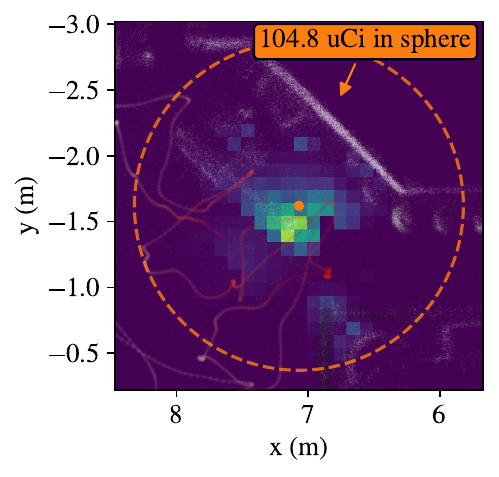} &
        \hspace*{-0.2cm}\includegraphics[height=0.3\textwidth]{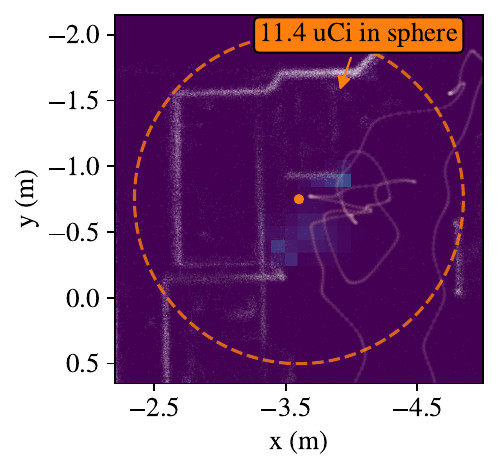} &
        \hspace*{-0.2cm}\includegraphics[height=0.3\textwidth]{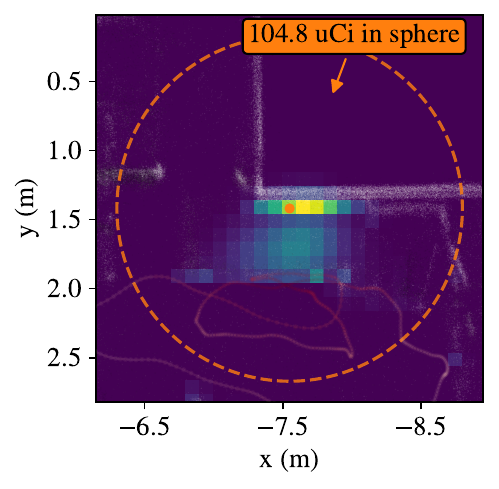} \\[-6pt]

        \multicolumn{3}{l}{\bf{d}} \\[-10pt]
        \multicolumn{3}{c}{\includegraphics[width=0.83\textwidth]{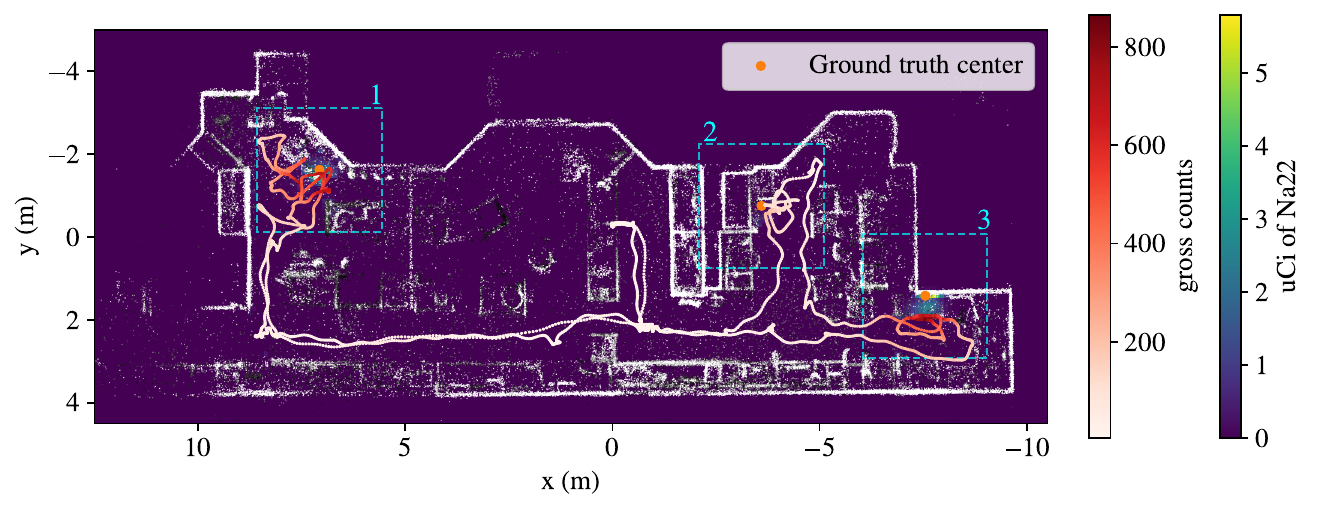}}\\[-5pt]

    \end{tabular}
    \ifexportfigs
    \else
    	\caption{{\bf Distributed source reconstruction results}. ({\bf a}) Setup of the three distributed source distributions placed in a cluttered laboratory environment. ({\bf b}) Cropped renderings of the 3D point clouds at the location of the three source distributions in A, colorized by the quantified Compton MAP-EM reconstruction results with the bottom 1\% of the reconstructed activity clipped to increase contrast with the point cloud. ({\bf c}) Top-down projections of the reconstruction in the three source areas, with a 1.25~m sphere (in orange) around the center of the ground-truth sources to determine the reconstructed source activity. ({\bf d}) Full top-down projection of the reconstruction result.}
   \fi
    \label{fig:distributed_source_results}
\end{figure}

The measurement took place in an active indoor laboratory environment wherein three distributed sources, separated by at least 5~m, were simultaneously staged.
The distributions were constructed with the arrayed point-sources in distinct configurations to model radioactive spills and are shown in Fig.~\ref{fig:distributed_source_results}a.
The first configuration contains a rectangular patch of sources ($6 \times 12$) on top of a table as well as a square patch of sources ($6 \times 6$) on the floor next to the table, with a total activity of 108~\textmu Ci \textsuperscript{22}Na.
The second configuration was a single smaller rectangular patch ($6 \times 4$) on the ground next to a fume hood, with a total activity of 24~\textmu Ci \textsuperscript{22}Na.
The final configuration was an $\mathsf{L}$-shaped patch comprised of three square patches ($6 \times 6$), with two of the patches on a set of flat boxes and one patch up against a wall.
The total activity of the final configuration was 108~\textmu Ci \textsuperscript{22}Na.

The MiniPRISM system was hand-carried to survey each area of the laboratory in detail for 1--2~min to model an operation in which the operator had access to the real-time count-rate information.
The total measurement time was 5.6~min and the generated map of the laboratory was $\approx$160~m$^2$.
The distances of closest approach to the center of the three source distributions were 0.53, 0.63, 0.54~m, in order of their description above and in Fig.~\ref{fig:distributed_source_results}a.

Similar to the point-source scenario, the results were generated as a single reconstruction including all measurement data.
The reconstruction energy region-of-interest (ROI) was $511 \pm 50$~keV and a total of 39,183 singles and 4,379 doubles were recorded in the ROI over the entire measurement.
Other reconstruction parameters include the following: 10~cm voxels, 125,754~occupied voxels (2.7\% of total), 3,342~poses (10~Hz) and 10~MAP-EM iterations.
An occupancy threshold of 10 points was used in order to reduce the number of occupied voxels and thus decrease the reconstruction time.
A higher pose frequency of 10~Hz was used here in order to more accurately capture the higher-acceleration movements that can occur in hand-held operation compared to sUAS surveys.

The $L_{1/2}$ regularization coefficients for the active coded mask and Compton MAP-EM reconstructions were set to $2 \times 10^{-2}$ and $8 \times 10^{-3}$, respectively.
The coefficients were determined via a coarse grid search to better match the total activity reconstructions of the three sources compared to ground truth.
Regularization strength is relative to the loss in the imaging objective function and, as a result, the optimal coefficients are expected to vary with statistics and imaging modality.
The coefficients were determined specifically for this data and thus may not extend to other measurements, though more general and algorithmic optimizations of such factors are an active field of research\cite{reader_2020} and will be explored in future work.

The results of the list-mode Compton MAP-EM (reconstruction time of 100.2~s) are shown in Fig.~\ref{fig:distributed_source_results}b-d.
The reconstruction of the large activity distributions (a.1 and a.3) are spatially accurate, capturing the 3D nature of the distributions representing spills on the ground or up on a wall.
The total reconstructed activities were estimated by summing all voxels in a 1.25~m radius sphere around the center of the true distributions.
The radius was selected as the size at which all three source activity reconstructions plateaued (Supplementary Fig.~S2) and the ground-truth center point was used as a reference point to quantitatively compare the reconstruction with the ground-truth activity.
The reconstructed activities of a.1 and a.3 (104.8 and 104.8~\textmu Ci) are in excellent agreement with ground truth of 108~\textmu Ci.
The reconstruction of source a.2 is observed to be noisy and the activity estimate (11.4~\textmu Ci) underestimates the true source activity (24~\textmu Ci).
This is believed to be because of a general lack of signal in this area, due to the lower overall source strength and the larger distance of closest approach in the survey.
There is also activity attributed to the vertical surface of the cart that was slightly nearer to the survey trajectory.
This incorrect location of activity nearer to the detector further skewed lower the source activity estimate.

The active coded mask MAP-EM reconstruction is included in Supplementary Fig.~S3.
The reconstructions were similarly spatially accurate to the Compton results, though more diffuse, exemplifying the differences in imaging resolving power between active coded mask and Compton imaging with MiniPRISM.
The reconstructed activities for sources a.1--a.3 were 118.2, 6.3 and 112.0~\textmu Ci of \textsuperscript{22}Na, respectively, and the total reconstruction time was 140.8~s.

In contrast to the point-source results presented above, the distributed source reconstructions do not yet quantify uncertainty in the activity estimates due to spatial reconstruction error, however we ascribe a 9.4\% uncertainty in the absolute scale of the response function used for reconstruction (see Methods).
Future work could include exploring other approaches to provide uncertainty estimates in reconstructed activities, such as Markov Chain Monte Carlo and approximate highest posterior density credible regions in MAP estimates \cite{cai_2018}.
These approaches introduce additional computational challenges and thus further work is necessary to develop a near real-time uncertainty quantification estimate for distributed sources.

\section*{Discussion}

The results presented above demonstrate that the quantitative SDF approach is capable of accurately and precisely localizing point-sources and imaging distributed sources in both the spatial and activity domains.
The images can be used to quickly find point-sources or map distributed sources in wide-area or cluttered environments as well as provide detailed contextual information to the user in order to quickly plan the next set of actions.
Furthermore, the methods can be executed on the timescale of seconds to minutes, allowing results to provide actionable feedback during the survey.
Additionally, the quantified images can be used to compute dose-rate maps in areas of interest surrounding the source which is critical in contamination assessment or dose-minimizing activities.
While this work focused on gamma-ray reconstruction, similar approaches can be implemented to enable quantitative SDF for neutron sources.


All SLAM-based approaches will be limited by the uncertainties in the real-time pose estimates.
While on the order of cm in most cases, the uncertainties may become significant in near-field imaging applications or in cases where high angular acceleration is applied to the system (e.g., fast twisting in hand-carried operations).
When coupled to systematic errors in the detector response, either from insufficient physics models or incorrect simulation geometries, the reconstructions can introduce unwanted image artifacts.
Additionally, some amount of solution degeneracy will also exist due to the finite angular resolution of the imaging system (e.g., 7--10 deg in MiniPRISM), the limited statistics in the data, and the potentially large number of degrees of freedom (i.e., voxels) in the model.
These effects can make the reconstruction susceptible to over-fitting and thus care must be taken to choose appropriate types and strengths of regularization.
Operator understanding, training, and real-time feedback can also help to mitigate over-fitting by ensuring sufficient spatial sampling and pose variation in important regions and visually inspecting that repeated passes do not substantially alter the solution.

The 511 and 662~keV energies reconstructed here represent a unique region in the energy spectrum where both active coded mask and Compton imaging have utility.
Below a few hundred~keV, the Compton imaging modality degrades as the number of measured doubles events diminishes, the separation between two-site interactions decreases (increasing the uncertainty in the scattering axis), and the effect of Doppler broadening (increasing the uncertainty in the scattering angle) becomes more pronounced \cite{du_2001}, in which case only active coded mask should be used.
Conversely, above several hundred~keV the number of photo-electric absorption-only events in a single 1-cm$^3$ CZT detector decreases significantly and Compton events become more likely, reducing the efficacy of the masking material, in which case only Compton imaging should be used.
Here we simply demonstrate the capability to use both modalities separately, but work has been done with mixed-event reconstruction \cite{lee_2007} in the energy regime where both modalities are still applicable and could be applied to SDF in future work.

Finally, the binary voxel-occupancy model (occupied and free) currently implemented is limited to reconstructing gamma-ray emissions to surfaces observed by the lidar.
A tri-state occupancy model to classify unknown space, in addition to free and occupied space, would enable the ability to reconstruct within enclosed volumes and other areas not directly seen by the lidar.
This capability will complement the work towards inclusion of attenuation in the physics model to account for both observed and possibly unknown attenuating material in the scene (e.g., material inside cargo containers in addition to the steel enclosure), enabling tomographic reconstruction.

\section*{Methods} \label{sec:methods}

This section describes the MiniPRISM free-moving gamma-ray imaging system, detector response characterization, image reconstruction algorithms, and the uncertainty quantification approach used in this work.

\subsection*{Detector System}

The MiniPRISM system is a free-moving multi-modal omnidirectional gamma-ray imager, consisting of a partially populated 3D grid (6$\times$6$\times$4) of 1-cm$^3$ coplanar-grid CZT detectors \cite{luke_1994, luke_1995}.
The 3D array of independently readout detector modules facilitates both the active coded mask and Compton imaging modalities, enabling omni-directional imaging of sources across a wide range of energies relevant to nuclear security, safety and safeguards (tens of keV to several MeV).
The detector system is paired with the Localization and Mapping Platform (LAMP) \cite{pavlovsky_2018} which includes the auxiliary sensor package (visual camera, lidar, and IMU) and single-board computer for real-time contextual and gamma-ray data processing.
The entire system is powered with a single 98 W-hr Li-ion battery, typically offering about one hour of operation on a single charge.
Battery life may fluctuate depending on the computational demand from image reconstruction and SLAM.
The system is platform agnostic, meaning it is capable of being hand-carried or mounted to a vehicle such as an automobile, ground robot, or sUAS.

\subsection*{Detector Response Characterization}

Quantitative gamma-ray emission reconstructions require characterization of the angular response of the detector system.
Here we characterize the MiniPRISM system response through experimentally benchmarked simulations.
The simulations were performed using the Monte Carlo particle-transport code Geant4 \cite{agostinelli_2003, allison_2006,allison_2016} and enabled parametric study of the angular response functions over the relevant emission energy domain.
The simulations track the interaction positions and energy depositions of source photons and secondary particles in the detector array.
Energy depositions are blurred according to a measured energy-dependent resolution function.
Particles are emitted as a uniform cone beam pointed at the detector, with an apex at the source position and an opening angle defined such that the cone fully encompasses the detector system.
Events of interest are tallied and used to compute a response $\eta$ in units of effective area.
For a particular source simulation (i.e., a single direction and energy), the effective area is computed using
\begin{equation}\label{effective_area_sim}
    \eta = \frac{2\pi r_s^2 N_d (1-\cos\theta)}{N_s}\,,
\end{equation}

\noindent where $N_s$ is the number of simulated particles, $r_s$ is the source-detector distance, $\theta$ is the cone opening angle, and $N_d$ is the number of tallied events in the detector.
For singles imaging, $N_d$ includes only the events that deposit energy within the specified energy ROI in a single detector module.
Therefore, in addition to energy and direction, $\eta$ is computed for each individual detector in the array.
It is the unique modulation pattern of $\eta$ across the detectors (for a given source direction and energy) that can be leveraged in reconstructing the source.
In Compton imaging, two-interaction event sequences (i.e., doubles) with a total energy deposition within the energy ROI are tallied across the entire detector array.
The result is an energy- and direction-dependent $\eta$ for the whole array.

Statistical uncertainties are included in the simulated $\eta$ to account for Poisson counting statistics (i.e., $\delta N_d = \sqrt{N_d}$).
Additional systematic uncertainties will inevitably exist for the simulated results, arising from model inaccuracies such as geometry, material composition and gamma-ray detection physics (e.g., charge transport/collection in CZT).
For example, a small dead layer beneath the anode is a well-documented phenomenon \cite{he_2001} in coplanar-grid CZT with a size roughly the same as the grid pitch.
Here we include a 1.25~mm dead layer in simulation to accommodate the 1.05~mm inner grid pitch as well as the larger grid lines near the edges \cite{luke_1997}.
We also use a 225~\textmu m surface cut on the remaining five sides of the cubes to account for reduced sensitivity due to electric field non-uniformities from the surface and from other nearby detector modules/electronics.
The value of this cut was set empirically to better match the measured response.
Additional systematic uncertainties are not explicitly accounted for in the calculation of $\eta$, though are discussed below.

Benchmark measurements were performed with a $4\pi$ detector scanning system designed and built at LBNL.
The scanning system consists of a rotatable platform, on which the detector system is placed, and a rotatable arm, on which the radiation source is attached (Supplementary Fig.~S4).
The combination of the platform rotation around the azimuth and the arm rotation along the polar direction facilitates a source position anywhere in the 4$\pi$ detector coordinate system.
The source holder arm is extendable, allowing for the source to be placed at a varying distances away from the system ranging from 44--143~cm.

Several independent scans were performed with relatively strong laboratory-scale sources (hundreds of \textmu Ci), with gamma-ray energies ranging from 60 to 1332 keV.
Source directions were nominally defined with the Hierarchical Equal Area isoLatitude Pixelation of a sphere (HEALPix) library \cite{gorski_2005} and the lowest resolution partition parameter $N_\textrm{side} = 1$.
The $N_\textrm{side} = 1$ discretization describes four-point azimuthal scans at polar angles of 49.2, 90, and 131.8~deg (12 total positions).
The 49.2 and 131.8~deg scans include azimuthal angles at 45, 135, 225 and 315~deg.
The equatorial (90~deg) scan includes azimuthal angles at 0, 90, 180 and 270, aligning with the symmetry axes of the detector system.
The reader is encouraged to refer to Figure 4 of G\'orski et al.\cite{gorski_2005} for a visual representation.
The source-detector distance was set at 110~cm and considered to be in the far-field ($>$10 times the roughly 7~cm spatial extent of the detector array).
The source remained static at each position for 5~min, resulting in a total individual scan time of 1~hr.
In contrast to equation~(\ref{effective_area_sim}), the sources were assumed to emit isotropically and thus the measured effective area was computed using
\begin{equation}\label{effective_area_meas}
    \eta = \frac{4 \pi r_s^2 N_d}{B t A e^{-\lambda T}}\,,
\end{equation}

\noindent where $B$ is the branching ratio of the decay of interest, $t$ is the dwell time, $A$ is the calibrated activity (decay rate) marked on the laboratory source, $\lambda$ is the decay constant and $T$ is the time elapsed since the source activity calibration.
Here the dwell time $t$ is much less than the half-life ($\tau_{1/2} = \log(2)\cdot\lambda^{-1}$) of the measured sources.
A background subtraction was performed with a linear fit between the average counts in neighboring windows on both sides of the spectral region-of-interest.
Neighboring window widths varied across the measured source energies, as some sources (e.g., \textsuperscript{133}Ba) have closely-separated gamma-ray emission lines and care was taken to not include source photons in the background estimation.

The largest sources of uncertainty in the calculation of $\eta$ arise from uncertainty in $A$, $r_s$ and $N_d$.
The laboratory sources were calibrated by the manufacturer and reported to have a 3\% total uncertainty (99\% confidence).
The MiniPRISM system was placed on the scanning system and aligned by eye to be in the center of the rotating platform and in line with the height of the rotating arm.
Given the potential for misalignment in two joints of the system, a conservative uncertainty of $r_s = 110.0 \pm 2.5$~cm was assumed.
Poisson counting statistics are included for the uncertainty on $N_d$ as $\sqrt{N_d}$.
Finally, the rotating arm is known to exhibit varying degrees of sag along the polar angle due to the weight and extension of the arm away from the pivot point.
This effect is pronounced at source directions that fall in line with one of the axes of symmetry of the detector array, where outer detectors should perfectly occlude inner detectors.
With a small amount of source-arm sag, additional streaming paths become available to many of the previously occluded detectors and the detection sensitivity can increase significantly, especially at lower energies.
To mitigate the discrepancy between simulation and measurement, the simulations were performed with a 5-deg sag in the polar angle for source directions around the equatorial plane (HEALPix indices 4--7 for $N_\textrm{side}=1$).

\begin{figure}[htb!]
    \centering
    \hspace*{-0.25cm}\includegraphics[width=0.95\textwidth]{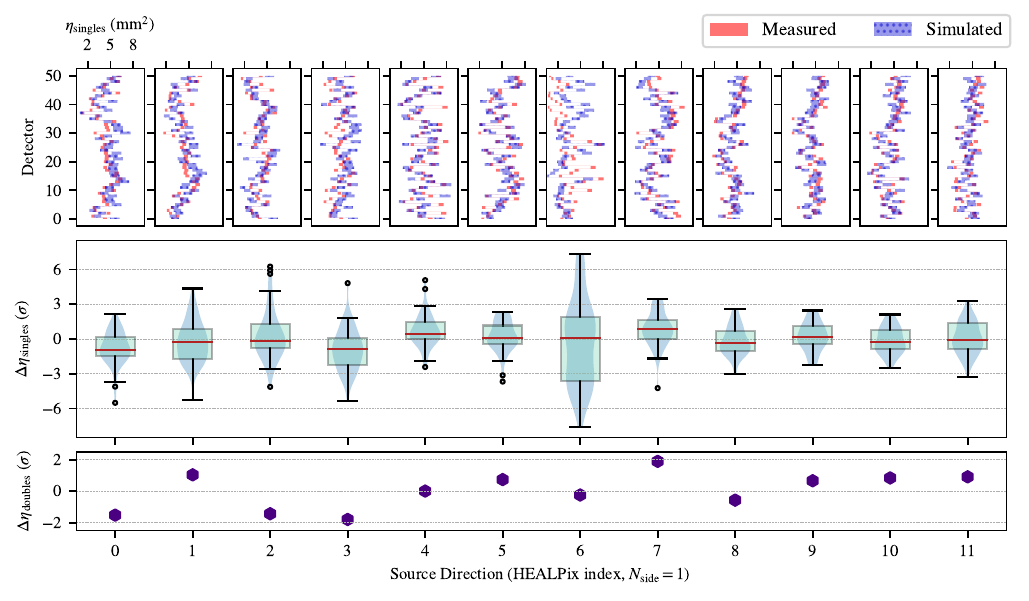}\\[-7pt]
    \ifexportfigs
    \else
    	\caption{
     	  {\bf Detector response characterization}.
     	   MiniPRISM far-field 662-keV gamma-ray detector response characterization (singles and doubles) at 12 source directions around 4$\pi$ defined by the HEALPix discretization scheme with $N_\textrm{side} = 1$.
     	   Measurements were performed with the scanning system described in the text, using a 182.5~\textmu Ci \textsuperscript{137}Cs at 110~cm from the center of the detector array and 5-min acquisitions at each source direction.
     	   Simulations were performed using Geant4.
     	   The top pane shows the measured (red) and simulated (blue) full-energy singles effective area (mm$^2$) of individual detector elements, with shading denoting uncertainties (discussed in the text), across the 12 source directions.
     	   The middle pane shows a box-and-whisker plot of the per-detector effective area differences between measurement and simulation in the top plot, expressed in units of Gaussian $z$-scores ($\sigma$).
      	  The boxes and whiskers describe the interquartile range (IQR) and 1.5$\times$ICR, respectively, and points indicate outliers outside the extent of the whiskers.
      	  A red line is drawn in each box to denote the median of the data.
      	  Finally, a violin plot is overlaid to provide a sense of the distribution (via a 1D kernel density estimator) of the data across the vertical dimension.
      	  The bottom pane shows the doubles effective area differences between measurement and simulation of the entire detector array, in units of Gaussian $z$-scores.
      	  }
   \fi
    \label{fig:response_characterization}
\end{figure}

Figure~\ref{fig:response_characterization} shows the singles and doubles effective area characterization results at 662~keV.
The top pane shows the per-detector agreement of the single effective area of measurement (red) and simulation (blue) across the 12~source directions, where uncertainties are represented by the extent of each rectangular marker.
The general trends and amplitudes are seen to qualitatively agree quite well.
The middle pane illustrates the deviations with a more quantitative approach, displaying the differences of the per-detector effective areas between measurement and simulation (in units of Gaussian $z$-scores) as standard box-plots (light green) across the 12 source directions.
Violin plots (light blue) are overlaid to provide a sense of the distribution of the data points along the vertical dimension.

The majority of simulations of individual detector efficiencies agree with the observed data to within the 1-2~$\sigma$, and the overall simulated detector system singles efficiency is within 1$\sigma$ for all measured angles, though some variance is observed for several source directions.
First, directions 0--3 are positioned at a polar angle of ~48$^\circ$, corresponding to a source above the system where the electronics and battery enclosure occlude the detector array.
The Geant4 model of the enclosure includes the battery, single board computer and analog-to-digital converter (ADC) boards, though does not have significant detail to include other potential occluding and scattering media such as electronic components, fasteners and cabling.
Therefore, it is expected that some level discrepancy exists for these source directions.
The remaining source directions agree to 1-2~$\sigma$ with much smaller variance, except for position 6 in which the lidar and camera are positioned directly in front of the detector array.
Specific details of the lidar and camera internals and material composition are unavailable in manufacturer documentation and thus the Geant4 models are only approximate, leading to a larger variance in the effective area comparison at this position.
Due to this finding, measurements with the MiniPRISM system are designed to reduce the time at which the lidar and camera package are placed between the suspected source location and the detector array.
While not explicitly modeled or characterized here, we assume a human operator may have a similar effect when hand-carrying the system, and thus care is taken to also reduce the time at which the operator is between the detector and source.
We acknowledge that this may not be possible in practice where potential source locations are not known \emph{a priori}, and thus additional work is needed to refine the response in these directions.
Finally, the bottom panel of Fig.~\ref{fig:response_characterization} shows the differences between the measured and simulated doubles effective area for the entire detector array in units of Gaussian $z$-scores.
Agreement is made for all source directions to $\leq 2\sigma$.

\begin{figure}[htb!]
    \centering
    \begin{tabular}{cc}
        \multicolumn{1}{l}{\bf{a}} & \multicolumn{1}{l}{\bf{b}} \\[-3pt]
        \hspace*{-0.25cm}\includegraphics[width=0.46\columnwidth]{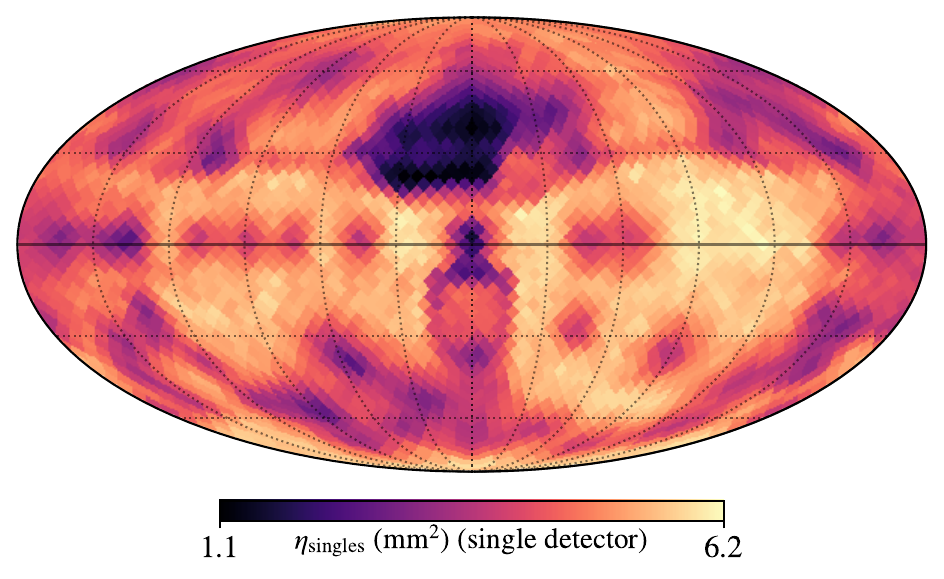} &
        \hspace*{-0.25cm}\includegraphics[width=0.46\columnwidth]{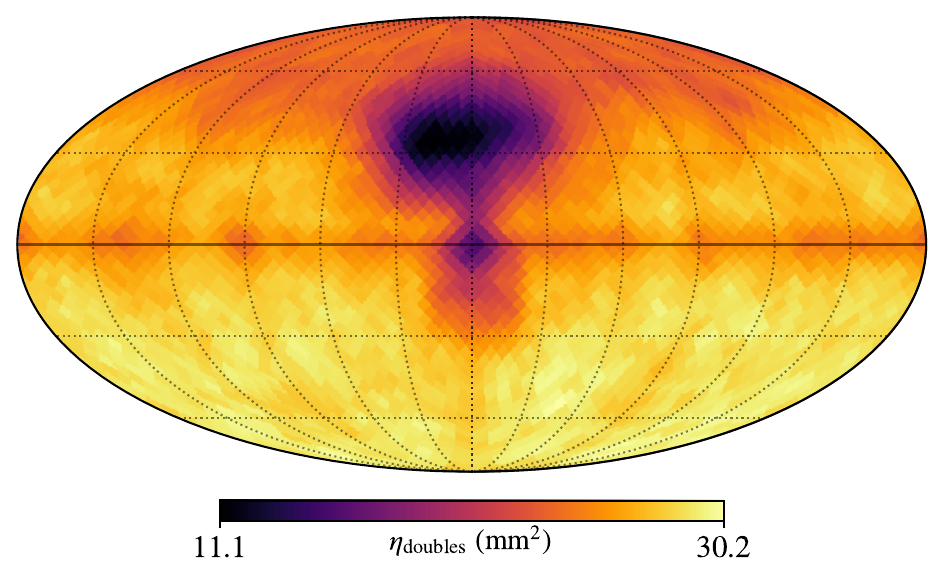}\\ [-5pt]
    \end{tabular}
    \ifexportfigs
    \else
    	\caption{{\bf Simulated detector response examples}. ({\bf a}) 4$\pi$ singles effective area of a single detector near the center of the array (662~keV). ({\bf b}) 4$\pi$ doubles effective area of the entire detector array (662~keV). Both images use the Mollweide projection with the origin at the center of the detector array and the positive X and Z axes pointing through the camera/lidar package and electronics box, respectively. Graticules are shown every 30~deg in polar and azimuth.}
   \fi
    \label{fig:miniprism_model_response}
\end{figure}

In this work, quantitative SDF is first demonstrated for sources with gamma-ray energies of, or near, 662~keV.
Therefore, the characterization of detector response at energies other than 662~keV will not be presented here.
Following characterization, higher-resolution simulations ($N_\textrm{side}=16$ or 3072 source positions) were performed at source emission energies relevant to the considered sources in this work (i.e., 511 and 662~keV).
Example simulated responses are shown in Fig.~\ref{fig:miniprism_model_response} for singles (one detector) and doubles at 662~keV.
Distinct anisotropy is observed due to occlusions from the auxiliary sensors, electronics and battery as well as self-occlusions from other detector modules in the array.

\subsection*{Imaging Modalities}

In addition to the gamma-ray source localization and mapping capabilities provided by signal modulation due to the motion of the system near an area of interest (i.e., proximity mapping), this work reconstructs gamma-ray images with the active coded mask and Compton imaging modalities.
The active coded mask modality relies on the energy- and direction-dependent modulation of incident gamma-ray flux from both passive materials (e.g., battery, contextual sensors) as well as active materials (i.e., other detector elements) in the system \cite{ziock_2008,schultz_2009}.
Active elements acting as both detector and occluding material decreases the overall weight of the system by removing the need for heavy Pb or W masks, increases the detection sensitivity and imaging field-of-view, and can be used to discriminate full-energy and down-scattered absorption events \cite{johansson_1980}.
Individual detector modules are arranged such that the spatial pattern of events in the array uniquely (ideally) determines the direction of the gamma-ray source.
However, as the energy and penetrative power of the gamma-rays increase (above a few hundred~keV), the uniqueness of the full-energy signal modulation degrades and the likelihood of scattering events increases.

The design of the active detector array facilitates Compton imaging of gamma rays that scatter in one detector and are absorbed in another.
Higher-order scattering sequences are excluded in this work as the analysis is significantly more complex and the Compton events are dominated ($\gtrsim$90\%) by doubles events.
The positions and energy depositions of the two interactions define the axis and opening angle of a cone that includes the direction of the incident gamma ray.
The overlap of multiple cones in 3D reveals the source location.

Compton kinematics dictate that the cone generated by the two-interaction event sequence has a vertex at the first interaction location.
In some imaging systems, the event time resolution and distance between interactions are such that time-of-flight (TOF) methods can be used to determine sequence ordering ($a \rightarrow b$ or $b \rightarrow a$) \cite{boggs_2000}.
For MiniPRISM, the distance between interactions is on the order of centimeters and the timing resolution is $\sim$2.4~\textmu s, therefore TOF cannot be used.
If $E \leq 256$~keV, kinematics assert that the first interaction will always deposit less energy than the second.
However, the energy regime in which Compton imaging is primarily used here is $>256$~keV and thus the sequence ordering can be ambiguous.
Previous work has taken a single cone approach, assuming the higher energy deposition event occurred first \cite{lehner_2004,lambropoulos_2010}.
To maximize sensitivity with limited statistics, our approach is to always project both possible cones with a simplified weighting for each cone based on the Klein-Nishina scattering and photoelectric absorption cross-sections.
The weighting is similar to the event selection criterion used in \cite{bandstra_2011}, though in that case still only one cone was projected.

In both active coded mask and Compton imaging, energy selection of photo-peak events is employed to image gamma rays from unique nuclear decay lines of specific radioisotopes of interest and to isolate source events from background.
Furthermore, as our reconstruction approach is performed in a globally consistent adaptive image space, we can perform data cuts in both space and/or time (e.g., selecting a portion of the trajectory and associated radiation data near a suspected source location) to further isolate source from background.

\subsection*{Scene Data Fusion}

In the two source measurement scenarios described in the Results section and the Supplementary Information, the data collection to image reconstruction workflow is as follows.
As the system is moved freely through an environment, the contextual sensors and SLAM solution provide real-time (up to 10~Hz) globally consistent estimates of the system trajectory (as a time series of position and orientation) and the surrounding 3D environment (as a point cloud).
At a given point in time, the trajectory extents (with a configurable padding) are used to define the bounds on which the point cloud is voxelized at some set resolution (e.g., 20~cm).
The generated voxel space retains the concept of occupancy -- i.e., whether a voxel contains a point (or a configurable minimum number of points) from the point cloud, and the occupancy grid defines the reconstruction space.

The time correlation of radiation data and the system trajectory can be performed in two ways.
As the radiation data is collected in list-mode (i.e., event-by-event), the system trajectory can be interpolated to the radiation event times, providing system pose estimates for each gamma-ray event or Compton sequence in the detector, facilitating list-mode type reconstruction algorithms.
The radiation data can also be binned according to the system trajectory times in order to run bin-mode type reconstructions, though there is inevitable information loss with this approach as quick translations or rotations (e.g., twisting a hand-carried system) may not be captured given the frequency at which the system pose is computed.
The nature of the data or the algorithm itself can sometimes dictate the type of reconstruction (i.e., bin-mode or list-mode).
In cases where either option is available, there are computational advantages to choosing one over the other -- e.g., bin-mode reconstruction times will scale as $P \times D$ where $P$ and $D$ are the number of poses and detectors, respectively, and list-mode will scale as the number of detected events, $N$.
Apart from high count-rate environments, typically $N < P \times D$ and thus a list-mode implementation is favorable.
In this work, the active coded mask PSL algorithm is implemented in a bin-mode reconstruction (due to current implementation constraints) while all other reconstruction algorithms (Compton PSL, active coded mask MAP-EM and Compton MAP-EM) are performed in list-mode.
The pose-correlated radiation interaction data and 3D occupancy grid are then sent to the reconstruction algorithm to produce an image (i.e., gamma-ray emission estimates on the 3D occupancy grid).

\subsection*{Image Reconstruction}

The objective of the imaging problem as defined in this context is, given some measured data $\boldsymbol{y}$, to form a non-negative image estimate consisting of $J$ voxels $\hat{\boldsymbol{w}} \in \mathbb{R}_+^{J}$, with units of gamma-ray emissions per unit volume per second, that satisfies the following
\begin{equation} \label{eq:objective}
    \hat{\boldsymbol{w}}, \hat{b} = \underset{\boldsymbol{w}, b}{\textrm{argmin}}[\ell (\boldsymbol{y} | \boldsymbol{w}, b) + \beta R(\boldsymbol{w})] \, ,
\end{equation}

\noindent where $b$ is a constant background rate, $R(\boldsymbol{w})$ is a regularization applied to the image, $\beta$ is a parameter to control the strength of the regularization, and $\ell$ is the negative logarithm of the Poisson likelihood
\begin{equation}\label{eq:poislikelihood}
    \ell(\boldsymbol{y} | \bar{\boldsymbol{y}})=[\bar{\boldsymbol{y}} - \boldsymbol{y} \odot \log{\bar{\boldsymbol{y}}}+\log[\Gamma(\boldsymbol{y}+1)]]^T \cdot \textbf{1} \,,
\end{equation}

\noindent where $\odot$ denotes element-wise multiplication, $\Gamma$ is the Gamma function, and $\bar{\boldsymbol{y}}$ is the mean estimate from our linear model $\bar{\boldsymbol{y}} = \boldsymbol{K} \cdot \hat{\boldsymbol{w}} + \hat{b} \boldsymbol{t}$, where $\boldsymbol{K}$ is the system matrix that maps image space to data space and $\boldsymbol{t}$ are the measurement dwell times.

Here $\boldsymbol{y}$ is acquired in list-mode and consists of $N$ events.
In the singles reconstruction paradigm (i.e., proximity and active coded mask), each $y_n$ is a single interaction event in detector $d$ at pose $p$ with energy deposition $E$.
The pose of the system represents the position $\vec{x}$ and orientation $\vec{\Omega}$ of the system in a global coordinate frame at time $t$ (provided by SLAM).
In the Compton reconstruction paradigm, each $y_n$ represents an event-pair (i.e., a Compton sequence) of two time-correlated interactions in detector pair $d_a$ and $d_b$ with energy depositions of $E_a$ and $E_b$, respectively, at pose $p$ with total energy deposition $E=E_a+E_b$, which is considered within the full-energy window of the energy of interest.
The two interactions are assumed to be a Compton scatter followed by a photoelectric absorption, though often we are unable to determine the sequence ordering ($a \rightarrow b$ or $b \rightarrow a$).

The $\{n,j\}$ element of the list-mode system matrix $\boldsymbol{K} \in \mathbb{R}_+^{N \times J}$ effectively represents the probability of a gamma-ray emitted from voxel $j$ producing event $y_n$.
In the case of singles imaging, this can be written as
\begin{equation}
    K_{nj} = \frac{\eta(\vec{r}_{nj}, \vec{\Omega}_n, E) \xi_{nj}}{4 \pi \rho(\vec{r}_{nj}, z)^2} \,,
\end{equation}

\noindent where $\eta(\vec{r}_{nj}, \vec{\Omega}_n, E)$ is the effective area of the detector at event~$n$ as seen from a source in voxel~$j$ with gamma-ray energy $E$, $\vec{r}_{nj}$ is the vector connecting voxel $j$ to the position of the detector at event $n$, $\vec{\Omega}_n$ is the orientation of the detector at event $n$, $\xi_{nj}$  is the transmission probability of a source photon travelling along $\vec{r}_{nj}$, and $\rho(\vec{r}_{nj}, z)$ is a regularized distance metric given by
\begin{equation}
    \frac{1}{\rho(\vec{r}_{nj}, z)^2} = \frac{|\vec{r}_{nj}|^2}{|\vec{r}_{nj}|^4 + z^4} \,,
\end{equation}

\noindent where $z$ is a nominal size of the detector system included here to mitigate unwanted near-field amplification and $|\cdot|$ is the Euclidean norm.
In the case of MiniPRISM, we set $z$ equal to 10~cm.
The transmission probability is always assumed to be unity here because of the additional computational burden of its calculation (i.e., ray-casting) and we have yet to employ real-time methods to estimate the material composition of intervening occupied voxels, though this is an active area of research \cite{bandstra_2021}.

In the Compton imaging case, $K_{nj}$ represents the intersection of both Compton cones associated with the $y_n$ event-pair with voxel $j$, where each cone is weighted by the Klein–Nishina and subsequent photoelectric cross-sections for the sequence as well as the Compton effective area of event-pair~$n$ from a source in voxel $j$ with total energy $E=E_a+E_b$.

To solve equation~(\ref{eq:objective}), a form of the iterative Expectation Maximization (EM) approach is used.
If no regularization is applied, the objective reduces to Maximum Likelihood (ML), in which traditional list-mode ML-EM \cite{parra_1998} update equations can be applied.
With regularization, we solve for the Maximum \emph{A Posteriori} (MAP) estimate using list-mode MAP-EM updates
\begin{equation}
    \hat{\boldsymbol{w}}^{l+1} = (\hat{\boldsymbol{w}}^l \oslash \boldsymbol{\psi}) \odot \boldsymbol{K}^T \cdot [\textbf{1} \oslash (\boldsymbol{K} \cdot \hat{\boldsymbol{w}}^l)] \,,
\end{equation}
\vspace{-15pt}
\begin{equation}
\boldsymbol{\psi} = \boldsymbol{\zeta} + \beta \boldsymbol{D}[R(\hat{\boldsymbol{w}}^l)]\,,
\end{equation}

\noindent where $\oslash$ denotes element-wise division, the sensitivity $\boldsymbol{\zeta} \in \mathbb{R}_+^J$ is the integrated response over all possible detected events (including those not measured in $\boldsymbol{y}$), and $\boldsymbol{D}$ is a vector of partial derivatives \cite{wilderman_1999}.
While fit in all bin-mode type reconstructions, background is currently not solved for in the list-mode formulation. The reconstructed weights will therefore be biased higher to account for background events in $y$, though in practice (and in the results presented here) we have observed this effect to be small. The inclusion of background fitting in list-mode reconstructions is left for future work.

Regularization is useful in mitigating over-fitting in the case of limited statistics and potentially large number of image voxels.
We have found good performance with sparsity regularization to force low intensity voxels to zero, effectively removing the voxels from the reconstruction.
Here we implement the $L_{1/2}$ regularization of the form used in Qian et al.\cite{qian_2011}
\begin{equation}
    R(\boldsymbol{w}) = \sqrt{\boldsymbol{w}} \cdot \boldsymbol{1} \, .
\end{equation}

Finally, the size of the system matrix will grow as the dimensionality of the problem increases and, in some cases, can become too large to store in computer memory, necessitating an on-the-fly computation of system matrix elements when needed.
However, parallelization of the forward (backward) projection operation is possible due to the independence of the voxels (events).
To maximize the gain of this parallelization, the reconstructions are executed on a GPU.
The implementation mitigates the memory burden as well as significantly increases computational performance, facilitating near real-time, order of seconds, reconstructions for most scenarios relevant in this application space.

\subsection*{Point-Source Localization}
Gamma-ray source localization and mapping using free-moving imagers is susceptible to bias and over-fitting due to variability in sensitivity across the space, limited angular resolution and the general ill-posedness of the problem.
Moreover, resultant spatial reconstruction errors can introduce large intensity estimation errors due to the inverse square falloff of counts with source-detector distance.
As a result, we developed a low-dimensional method called Point-Source Localization (PSL) \cite{hellfeld_2019_1,vavrek_2020} to improve spatial and intensity reconstruction in the point-source domain.
In this case, the source is constrained to a single image voxel and likelihood ratio tests are used to first compute source detection against a background-only model.
If a source is detected, additional likelihood ratio tests are used to compute spatial confidence intervals around the maximum likelihood source position.
The observed Fisher information matrix (i.e., the matrix of second derivatives of the negative log-likelihood function with respect to the point-source activity and background) is used to compute an approximate confidence interval for the source activity in each image voxel.
The union of the intervals inside a predefined spatial confidence interval (e.g., 2$\sigma$) is used to compute a conservative estimate of the confidence interval for the maximum likelihood source activity.
Additional details regarding uncertainty quantification can be found in Bandstra et al.\cite{bandstra_2021}.

\section*{Data Availability Statement}
The data used to generate the results will be made available on the \href{https://osf.io}{Open Science Framework} and access to the reconstruction software can be coordinated through the \href{https://ipo.lbl.gov}{LBNL Intellectual Property Office}.

\bibliography{ref}

\section*{Acknowledgements}

This material is based upon work supported by the Defense Threat Reduction Agency under HDTRA 10027-30529 and 13081-36239, and used the Lawrencium computational cluster resource provided by the IT Division at the Lawrence Berkeley National Laboratory (Supported by the Director, Office of Science, Office of Basic Energy Sciences, of the U.S. Department of Energy under Contract No. DE-AC02-05CH11231).
Distribution A: approved for public release, distribution is unlimited.
This support does not constitute an express or implied endorsement on the part of the United States Government.

\section*{Author contributions statement}

D.H., J.R.V., B.J.Q., and T.H.Y.J. curated the data. D.H. and T.H.Y.J. performed data analysis. D.H., M.S.B., D.L.G., and T.H.Y.J. developed the methodology. D.H., M.S.B., J.C.C., M.S., and T.H.Y.J. developed software. R.P., V.N., P.J.B. and J.W.C. contributed to the instrumentation. D.H. wrote the manuscript and all authors reviewed the manuscript.

\section*{Additional information}


\paragraph*{Competing interests}
The authors declare that they have no competing interests. The funders had no role in the design of the study; in the analyses or interpretation of data; in the writing of the manuscript, or in the decision to publish the results.

\clearpage

\section*{\centering Supplementary Information \\\smallskip Free-moving Quantitative Gamma-ray Imaging}
\begin{center}
Daniel~Hellfeld\textsuperscript{1,*},
Mark~S.~Bandstra\textsuperscript{1},
Jayson~R.~Vavrek\textsuperscript{1},
Donald~L.~Gunter\textsuperscript{2},
Joseph~C.~Curtis\textsuperscript{1},
Marco~Salathe\textsuperscript{1},
Ryan~Pavlovsky\textsuperscript{1},
Victor~Negut\textsuperscript{1},
Paul~J.~Barton\textsuperscript{1},
Joshua~W.~Cates\textsuperscript{1},
Brian~J.~Quiter\textsuperscript{1},
Reynold~J.~Cooper\textsuperscript{1},
Kai~Vetter\textsuperscript{1, 3},
Tenzing~H.\,Y.~Joshi\textsuperscript{1}\\
\medskip
\noindent \textsuperscript{1}Nuclear Science Division, Lawrence Berkeley National Laboratory, Berkeley, CA 94720 USA, \textsuperscript{2}Gunter Physics, Inc., Lisle, IL 60532 USA, \textsuperscript{3}Department of Nuclear Engineering, University of California, Berkeley, Berkeley, CA 94720 USA \\
\end{center}



\renewcommand{\thefigure}{S1}
\begin{figure*}[htb!]
    \centering
    \renewcommand{\arraystretch}{1}
    \setlength{\tabcolsep}{3pt}
    \begin{tabular}{cc}
        \multicolumn{2}{l}{\bf{a}} \\
        \multicolumn{2}{c}{\includegraphics[width=0.86\textwidth]{fig/container_stack_setup.pdf}} \\

        \multicolumn{1}{l}{\bf{b}} \\[-5pt]
        \multicolumn{2}{c}{\includegraphics[width=0.79\textwidth]{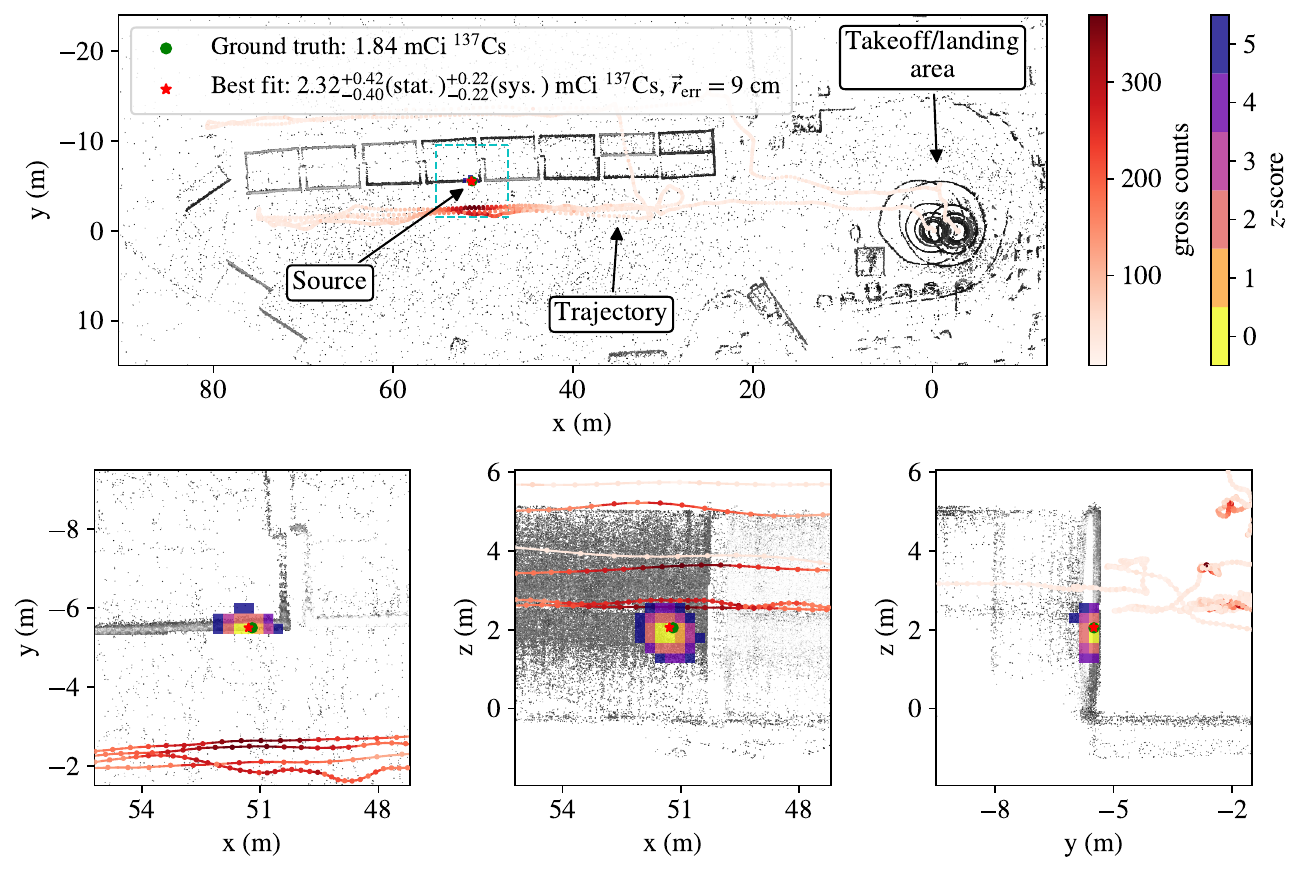}} \\[-5pt]

        \multicolumn{1}{l}{\bf{c}} \\[-5pt]
        \multicolumn{2}{c}{\fbox{\includegraphics[width=0.46\textwidth]{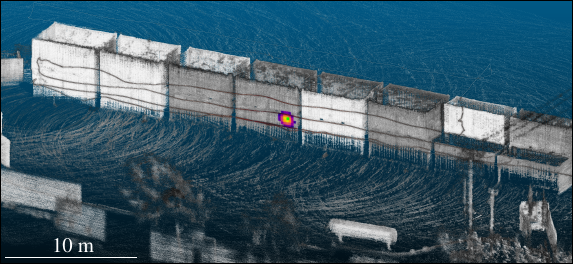}}}
    \end{tabular}
    \ifexportfigs
    \else
    	\caption{{\bf Point-source reconstruction results}. ({\bf a}) Point-source scenario in which a 1.84~mCi \textsuperscript{137}Cs source was placed $\approx$2.5~m above the ground on the exterior of a cargo container stack.
    	MiniPRISM was flown remotely on a sUAS and surveyed the stack in $\approx$7~min.
   	 ({\bf b}) Quantitative Compton PSL reconstruction following the survey of the cargo containers, including a full top-down projection as well as three zoomed-in projections near the true source location.
   	 ({\bf c}) Rendering of the colorized 3D point cloud.}
   \fi
    \label{fig:point_source_results_compton}
\end{figure*}
\renewcommand{\thefigure}{S2}
\begin{figure}[htb!]
    \centering
    \setlength{\tabcolsep}{1pt}
    \begin{tabular}{c}
        \multicolumn{1}{l}{\bf{a}} \\
        \includegraphics[width=0.75\textwidth]{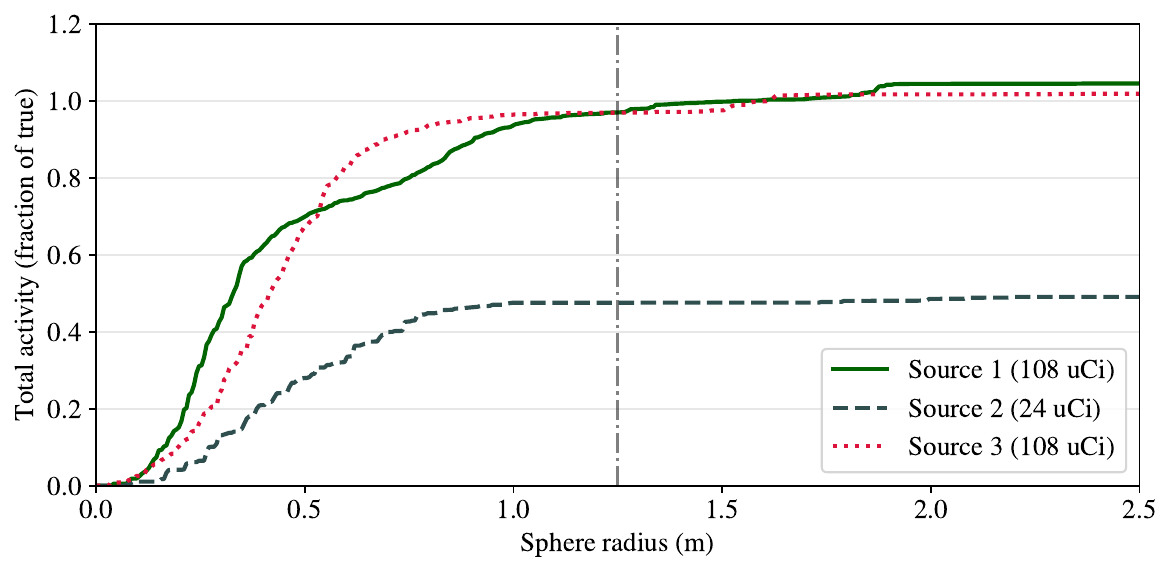} \\
        \multicolumn{1}{l}{\bf{b}} \\
        \includegraphics[width=0.75\textwidth]{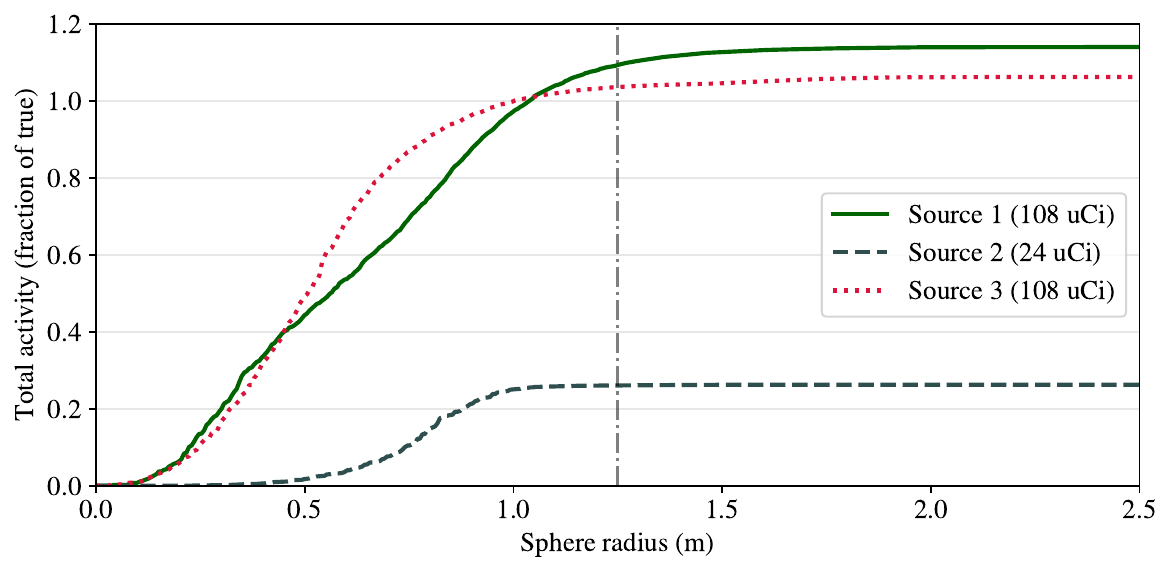}\\[-7pt]
    \end{tabular}
    \ifexportfigs
    \else
   	 \caption{{\bf Total reconstructed activity estimation for distributed sources}. Total reconstructed activity (as a fraction of the ground-truth activity) of the three source distributions in Fig.~3, calculated by summing the activity in voxels within a sphere around the center of the ground-truth source distributions.
   	The summed activities are computed as a function of the spherical radius and a radius of 1.25~m (shown by the gray dashed-dotted line) was selected as the point at which approximately all three source activity reconstructions plateaued.
   	Results are shown for Compton MAP-EM ({\bf a}) and active coded mask MAP-EM ({\bf b}) reconstructions.
   	 }
   \fi
    \label{fig:distributed_activity}
\end{figure}
\renewcommand{\thefigure}{S3}
\begin{figure}[htb!]
    \centering
    \setlength{\tabcolsep}{5pt}
    \begin{tabular}{*{3}{@{\hspace*{1mm}}c}}
        \multicolumn{1}{l}{{\bf a.1}} & \multicolumn{1}{l}{{\bf a.2}} & \multicolumn{1}{l}{{\bf a.3}} \\
        \fbox{\includegraphics[width=0.31\textwidth]{fig/distributed_source_1.pdf}} &
        \fbox{\includegraphics[width=0.31\textwidth]{fig/distributed_source_2.pdf}} &
        \fbox{\includegraphics[width=0.31\textwidth]{fig/distributed_source_3.pdf}} \\

        \multicolumn{1}{l}{\bf{b.1}} & \multicolumn{1}{l}{\bf{b.2}} & \multicolumn{1}{l}{\bf{b.3}} \\
        \fbox{\includegraphics[height=0.215\textwidth]{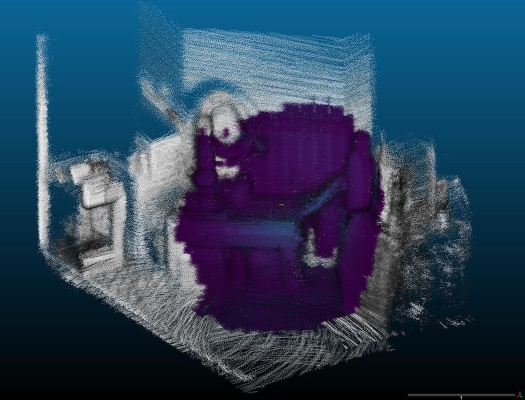}} &
        \fbox{\includegraphics[height=0.215\textwidth]{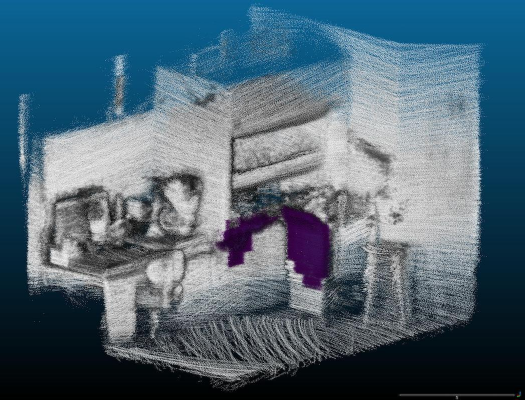}} &
        \fbox{\includegraphics[height=0.215\textwidth]{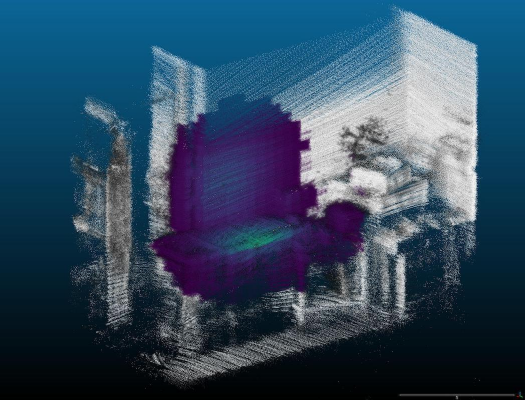}} \\

        \multicolumn{1}{l}{\bf{c.1}} & \multicolumn{1}{l}{\bf{c.2}} & \multicolumn{1}{l}{\bf{c.3}} \\[-3pt]
        \hspace*{-0.3cm}\includegraphics[height=0.3\textwidth]{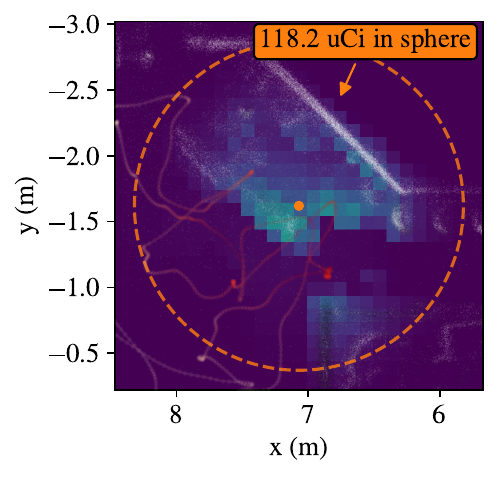} &
        \hspace*{-0.2cm}\includegraphics[height=0.3\textwidth]{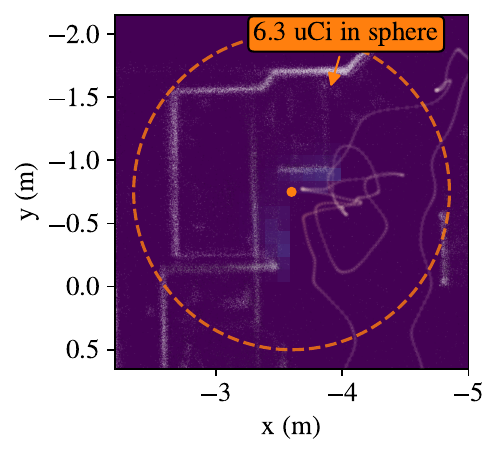} &
        \hspace*{-0.2cm}\includegraphics[height=0.3\textwidth]{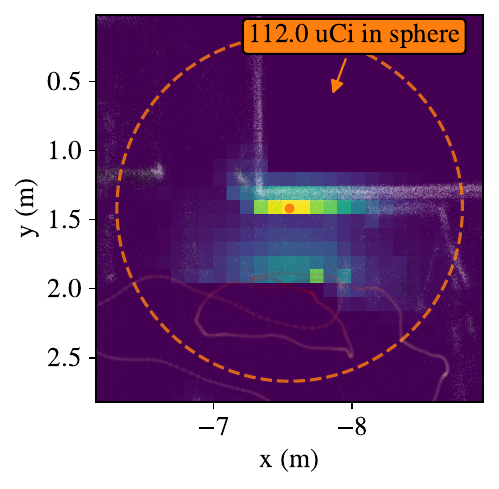} \\[-8pt]

        \multicolumn{2}{l}{\bf{d}} & \multicolumn{1}{l}{\bf{e}} \\[-2pt]
        \multicolumn{2}{l}{\hspace*{-0.5cm}\includegraphics[height=0.25\textwidth]{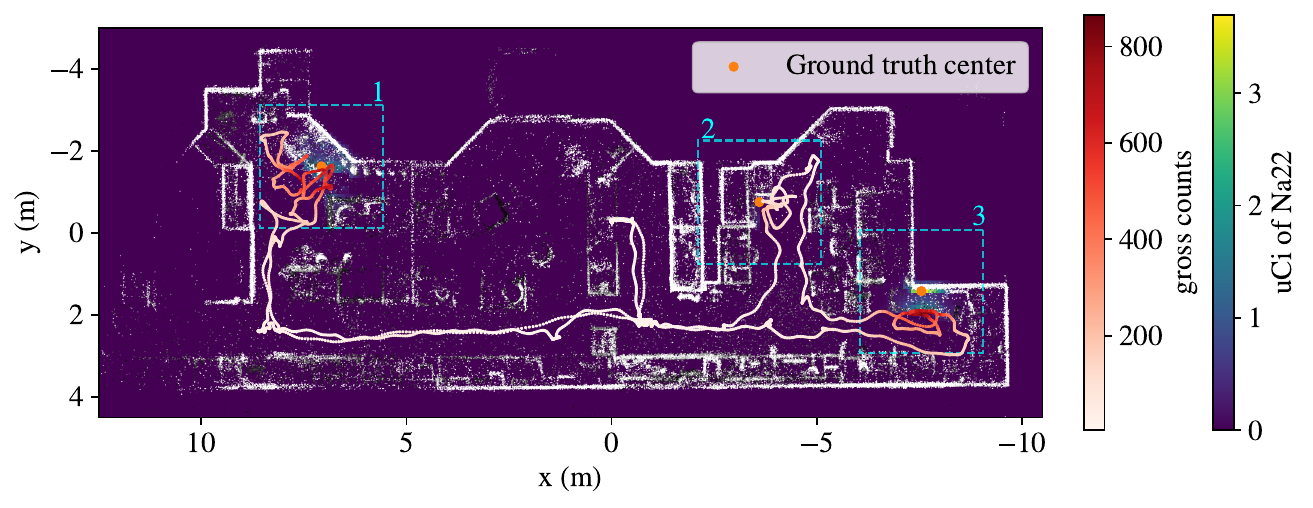}} &
        \multicolumn{1}{l}{\hspace*{0.0cm}\includegraphics[height=0.25\textwidth]{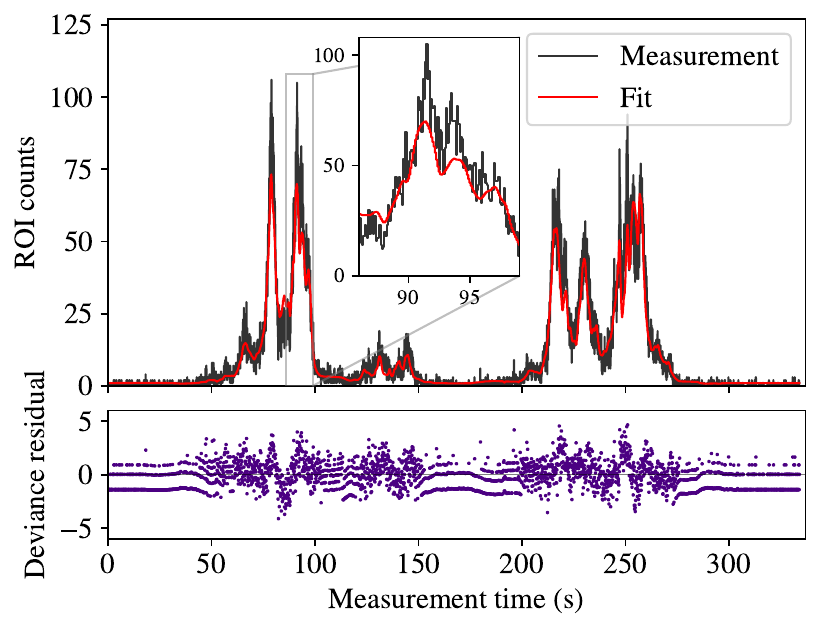}}\\[-6pt]

    \end{tabular}
    \ifexportfigs
    \else
    	\caption{
    	{\bf Distributed source reconstruction results}.
   	 ({\bf a}) Setup of the three distributed source distributions placed in a cluttered laboratory environment.
    	({\bf b}) Cropped renderings of the 3D point clouds at the source locations, colorized by the quantified active coded mask MAP-EM reconstruction results  with the bottom 1\% of the reconstructed activity clipped to increase contrast with the point cloud.
   	 ({\bf c}) Top-down projections of the reconstruction in the three source areas, with a 1.25~m sphere (in orange) around the center of the ground-truth sources to determine the reconstructed source activity.
    	({\bf d}) Full top-down projection of the reconstruction result.
    	({\bf e}) Forward projection of the reconstructed image overlaid on the measured data, summed across all detectors.}
   \fi
    \label{fig:distributed_source_results_occlusion}
\end{figure}
%
\renewcommand{\thefigure}{S4}
\begin{figure}[htb!]
    \centering
    \setlength{\tabcolsep}{5pt}
    \begin{tabular}{cc}
        \multicolumn{1}{l}{\bf{a}} & \multicolumn{1}{l}{\bf{b}} \\
        \fbox{\includegraphics[height=0.4\columnwidth]{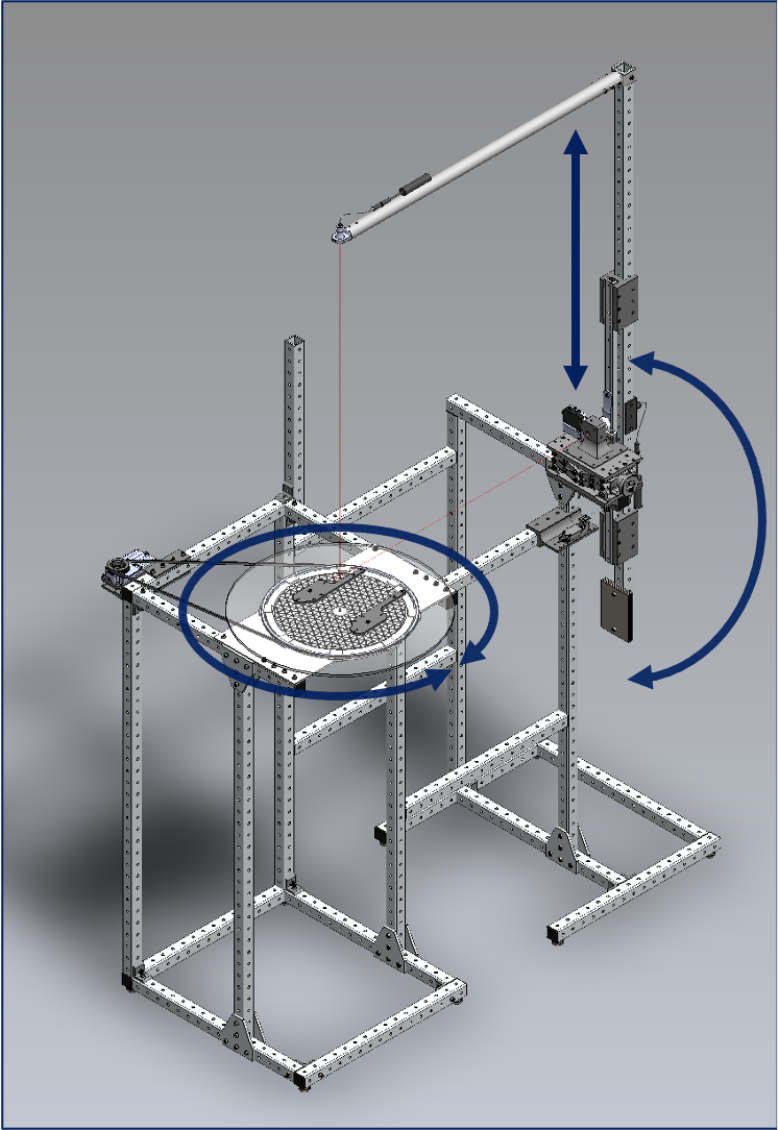}} &
        \fbox{\includegraphics[height=0.4\columnwidth]{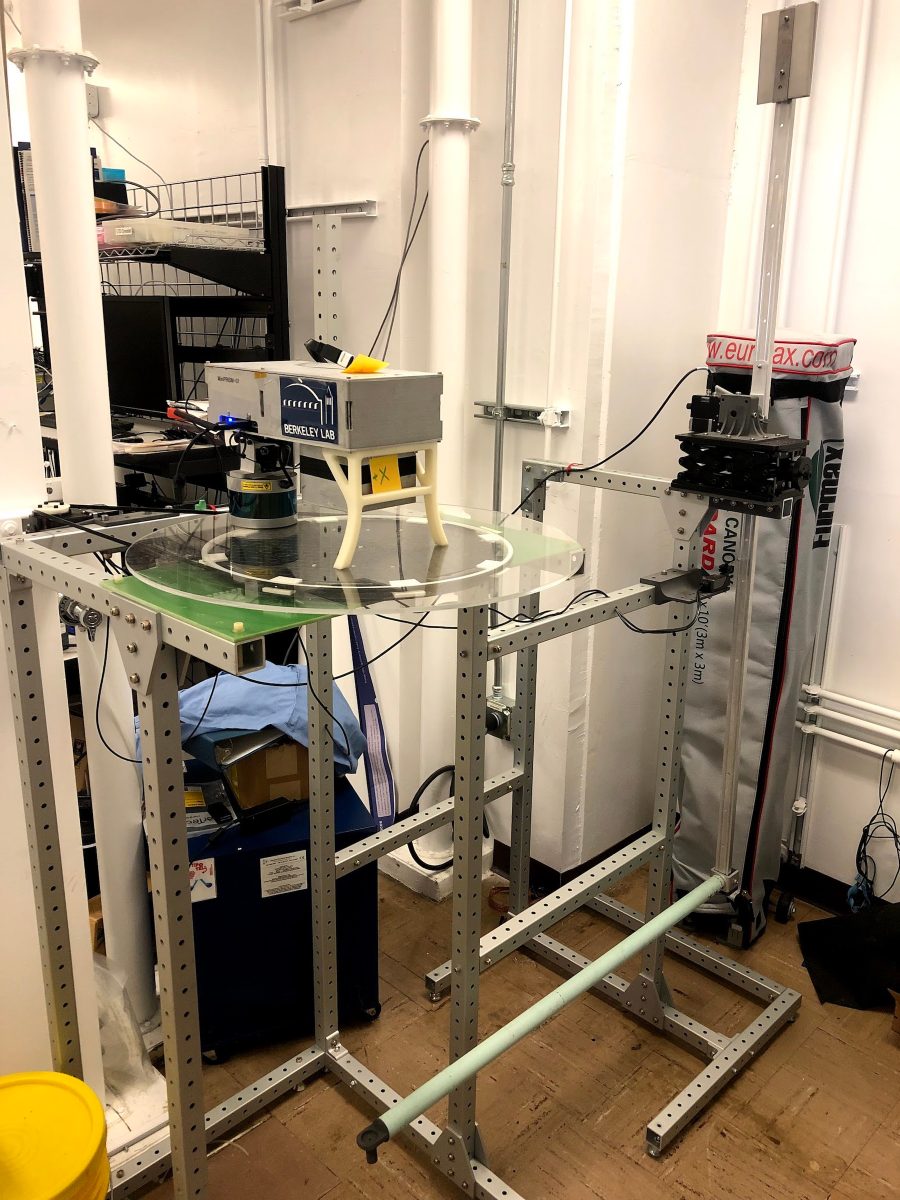}}
    \end{tabular}
    \ifexportfigs
    \else
   	\caption{{\bf Detector response characterization setup}. ({\bf a}) Rendering of the 4$\pi$ detector response scanner and concept of operation. ({\bf b}) The MiniPRISM system on the prototype scanner in the laboratory.}
   \fi
    \label{fig:miniprism_rig}
\end{figure}

\end{document}